\documentclass[twocolumn]{autart}    

\usepackage{graphicx}          

\usepackage{caption}
\usepackage{subcaption}

\usepackage{amsmath}
\usepackage{amsfonts}
\usepackage{amssymb}
\usepackage{cite}
\usepackage{color}  
\usepackage{algorithm} 
\usepackage{algpseudocode}

\def\rep#1{(\ref{#1})}

\newcommand{\R}{\mathbb{R}}

\def\send#1#2{\stackrel{#1}{\hbox to #2{\rightarrowfill}}}
\def\-{\!\!\!\!\!-}

 \def\qed{ \rule{.08in}{.08in}}
\def\eq#1{\begin{equation}#1\end{equation}}

\newcommand{\rank}{{\rm rank\;}}

\newcommand{\dfb}{\stackrel{\Delta}{=}}

\newtheorem{theorem}{Theorem}
\newtheorem{lemma}{Lemma}

\def\qed{ \rule{.1in}{.1in}}

\def\R{{\rm I\!R}} 

\newcounter{seqn}[equation]
\def\theseqn{\arabic{equation}\alph{seqn}}

\def\endseqn{\eqno \@seqnnum
$$\ignorespaces}
\def\@seqnnum{(\theseqn)}

\newskip\mcentering \mcentering=0pt plus 1000pt minus 1000pt

\def\meqalignno#1{
\halign to\displaywidth{
    \hbox to 0pt{\kern\displaywidth\llap{$##$}\hss}\tabskip=\mcentering
    &\hfil$\displaystyle{##}$\tabskip=\mcentering
   &&$\displaystyle{{}##}$\hfil\tabskip=\mcentering
    \crcr
    #1\crcr}}

\def\rep#1{(\ref{#1})}

\def\eq#1{\begin{equation}#1\end{equation}}

\def\dspace{\multiply\normalbaselineskip 150
          \divide\normalbaselineskip 100 \normalbaselines
          \csname @@normalbaselineskip\endcsname\normalbaselineskip}
\def\sspace{\multiply\normalbaselineskip 200
         \divide\normalbaselineskip 300 \normalbaselines
         \csname @@normalbaselineskip\endcsname\normalbaselineskip}
\def\sdspace{\multiply\normalbaselineskip 160
         \divide\normalbaselineskip 150 \normalbaselines
         \csname @@normalbaselineskip\endcsname\normalbaselineskip}

\def\df{ \; \stackrel{\triangle}{=} \; }

\def\@{\tilde}

\def\3dot#1{\buildrel\textstyle...\over#1}

\begin{document}

\begin{frontmatter}

\title{A Hybrid Observer for Estimating the State of a Distributed Linear System\thanksref{footnoteinfo}} 

\thanks[footnoteinfo]{This paper was not presented at any IFAC 
meeting. Portions of this paper were presented, in abbreviated form and without proofs,
  at the 2017 IEEE Conference on Decision and Control \cite{CDC17.1}. Corresponding author: L.~Wang. 
}

\author[uci]{Lili Wang}\ead{lili.wang.zj@gmail.com},    
\author[sb]{Ji Liu}\ead{ji.liu@stonybrook.edu},               
\author[yale]{A. Stephen Morse}\ead{as.morse@yale.edu}  

\address[uci]{Samueli School of Engineering,
University of California, Irvine}  
\address[sb]{Department of Electrical and Computer Engineering, Stony Brook University}        

\address[yale]{Department of Electrical Engineering, Yale University}             
          

\begin{abstract}  
A hybrid  observer is described for  estimating the state of an $m>0$ channel, $n$-dimensional,
continuous-time, linear system of the form $\dot{x} = Ax,\;y_i =
C_ix,\;i\in\{1,2,\ldots, m\}$. The system's state $x$ is
simultaneously estimated  by $m$ agents assuming each agent $i$
senses $y_i$ and receives appropriately defined data from each of
its current neighbors. Neighbor relations are characterized by a
time-varying directed  graph $\mathbb{N}(t)$
 whose vertices correspond to agents and whose arcs depict neighbor relations.
Agent $i$  updates its   estimate  $x_i$  of $x$ at ``event times''
$t_{i1},t_{i2},t_{i3},\ldots $
 using a local continuous-time linear observer and a  local parameter estimator which
  iterates $q$ times during each event  time interval $[t_{i(s-1)},t_{is}),\;s\geq1$
  to obtain an estimate of $x(t_{is})$.
Subject to  the assumptions that none of the $C_i$'s are zero, the
neighbor graph $\mathbb{N}(t)$ is strongly connected for all time,
and the system whose state is to be estimated is jointly observable,
it is shown that for any number $\lambda >0$, it is
 possible to choose $q$ and the local observer gains so that each estimate $x_i$
   converges to $x$  at least as fast as $e^{-\lambda t}$ does. This result holds whether or not agents
communicate synchronously, although in the asynchronous case it is
necessary to assume that $\mathbb{N}(t)$
 changes   in a  suitably defined sense.  Exponential convergence is also assured  if the event
  time sequences   of the $m$ agents are  slightly different than each other, although  in this case only if the
  system being observed is exponentially stable; this limitation
   however, is primarily a robustness issue shared by all  state estimators,
   centralized or not, which are operating in
    ``open loop'' in the face of small modeling errors.
The result also holds facing  abrupt changes in the number of vertices and arcs in the inter-agent communication graph upon which the algorithm depends.
\end{abstract}

\end{frontmatter}

\section{Introduction}

In \cite{tac.17} a distributed   observer is described for
estimating the state of an $m>0$ channel, $n$-dimensional,
continuous-time, jointly observable linear system of the form
$\dot{x} = Ax,\;y_i = C_ix,\;i\in \{1,2,\ldots, m\}$. The state
$x\in\R^n$ is   simultaneously estimated by $m$ agents assuming that
each agent $i$ senses  $y_i$ and receives the state  of each of its
neighbors' estimates. An attractive feature of the  observer
described in \cite{tac.17} is that it is able to generate an
asymptotically correct estimate of $x$ exponentially fast  at a
pre-assigned rate, if each agent's set of neighbors do not change
with time and the neighbor graph characterizing neighbor relations
is strongly connected. However, a shortcoming of the observer in
\cite{tac.17} is that it is unable to function correctly if the
network changes with time. Changing neighbor graphs will typically
occur if the agents are mobile. A second shortcoming of the observer
described in \cite{tac.17}  is that it is ``fragile'' by which we
mean that the observer is not able to cope with the situation when
there is an arbitrary abrupt change in the topology of the neighbor
graph such as the loss or addition of a vertex or an arc. For
example, if because of a component failure, a loss of battery power,
or some other reasons, an agent drops out of the network, what
remains of the overall observer will typically  not be able to
perform correctly and may  become unstable, even if joint
observability is not lost and what remains of the neighbor graph is
still strongly connected.

\vspace{-0.1in}

This paper breaks new ground by introducing a hybrid distributed
observer which overcomes the aforementioned difficulties without
making restrictive assumptions. To the best of our knowledge, this
observer   is the first provably correct  distributed algorithm
capable of generating  an asymptotically correct  estimate of a
jointly observable linear system's state in the presence of a
neighbor graph which changes with time under reasonably general
assumptions.  Although the observer is developed for continuous-time
systems,  it can very  easily be modified in the obvious way
  to deal with discrete-time systems.
  
  \vspace{-0.1in}

{\bf{Notation:}}
Given a collection of $n\times n$ matrices, $A_1,\; A_2,\;\ldots,\;A_m$, let $\text{diagonal}\{A_1,A_2,\ldots,A_m\}$ be the block diagonal matrix with $A_k$ as its $k$th diagonal block.
Given a collection of $n\times 1$ vectors, $v_1,\;v_2;\;\ldots, \;v_m$, let 
 $\text{column}\{v_1,v_2,\ldots,v_\}$ be the stacked vector  with $v_k$ as its $v$th sub-vector.
 For  an $n\times n$  matrix $A$, we let $\text{image} A$ denote the linear subspace spanned by matrix $A$.
 For two  $n\times n$  matrix $A_1$, and $A_2$, we let  $\text{image} A_2 \cap \text{image}A_2$ denote the intersection of the two images.

\vspace{-0.1in}
\subsection{The Problem}\label{prob}
\vspace{-0.1in}
We are interested in a   network of $m>0$ autonomous agents labeled
$1,2,\ldots, m$  which are able to receive information from their
``neighbors'' where by the  {\em neighbor} of agent $i$   is meant
any
 agent  who is in agent $i$'s reception range.
 We write
$\mathcal{N}_i(t)$ for the set of labels of agent $i$'s neighbors at
real \{continuous\} time   $t $   and always take agent $i$ to be a
neighbor of itself.
 Neighbor relations at time $t$  are
characterized  by a directed graph $\mathbb{N}(t)$ with $m$ vertices
and a set of arcs defined so that there is an arc from vertex $j$ to
vertex $i$  whenever agent $j$ is a   neighbor of agent $i$. Since
each agent $i$ is always a neighbor of itself, $\mathbb{N}(t)$ has a
self-arc at each of its vertices. Each agent $i$ can sense a
continuous-time signal
$y_i\in\R^{s_i},\;i\in\mathbf{m}\dfb\{1,2,\ldots, m\}$, where
\begin{eqnarray}y_i &=&C_ix,\;\;\;i\in
  \mathbf{m}\label{sys1}\\\dot{x} &= &Ax\label{sys2}
\end{eqnarray}
and $x\in\R^n$. It is assumed throughout that  the system defined by
\rep{sys1} and \rep{sys2} is \textit{ jointly observable}; i.e.,
  with $C = [C_1'\; C_2' \;\cdots \;C_m']'$, the matrix pair $(C,A)$ is observable. For simplicity, it is further assumed that
$C_i \neq 0,\;i\in\mathbf{m}$; generalization to deal with the case
when this assumption does
 not hold is straight forward.
The problem of interest is to develop ``private estimators'', one
for each agent, which, under ideal conditions without modeling or
synchronization errors,  enable each agent to obtain an
  estimate of $x$ which converges to $x$ exponentially fast at a pre-assigned rate .



 \vspace{-0.1in}
\subsection{Background}
\vspace{-0.1in}
The distributed state estimation problem has been under study in one
form or another for years. The  problem has been  widely studied as
\textit{a distributed Kalman filter problem}
\cite{Olfati-Saber2005,Olfati-Saber2007,Olfati-Saber2009,Khan2010,Khan2011,Kim2016,
Wu2016, Olfati-Saber2012}. A form of distributed Kalman filtering
is introduced in \cite{Olfati-Saber2005} for
 discrete-time linear systems; the underlying idea is to switch back and forth between conventional state
  estimation and a data fusion computation.
This approach is extended to continuous-time systems in
\cite{Olfati-Saber2007}. There are two  key limitations of the ideas
presented in  \cite{Olfati-Saber2005,Olfati-Saber2007}.   First,
 it is   implicitly assumed in each paper  that
data fusion \{i.e., consensus\} can be attained in finite time.
Second, it is also implicitly assumed that each pair $(C_i,A)$ is
observable; this restrictive assumption is needed in order to
guarantee that each local
 error covariance matrix  Riccati  equation
has a solution. Both papers also include  assumptions about graph
connectivity and information exchange which are more restrictive
than they need be.

\vspace{-0.1in}

 Discrete-time distributed   observers have recently appeared in \cite{Doos2013,Khan2014,Park2012,Park2012a,martins, MitraPurdue2016,  Acikmese2014,Ugrinovskii2013,DELNOZAL2019213,Rego2021}. 
None of these estimators admit
continuous-time extensions. 
 The algorithm in \cite{Doos2013}
works for fixed graphs with a relatively complicated topology design by studying the roles of each agent in the network.
The distributed observer proposed in   \cite{Khan2014}  can track the system only if the so-called Scalar Tracking Capacity condition is satisfied. 
Noteworthy among these is the  paper
\cite{martins} which  described a discrete-time linear system which
solves the estimation problem for jointly observable,  discrete-time
systems with fixed neighbor graphs  assuming only that the neighbor
graph
 is directed and
strongly connected. This is done  by recasting the estimation
problem as a classical decentralized control problem
\cite{wangdavison,corfmat}.
 Although   these  observers  are  limited to
  to  discrete-time systems, it has proved possible to make use of the ideas in
   \cite{martins} to obtain a distributed observer for
 continuous-time systems \cite{tac.17}.
In particular, \cite{tac.17} explains  how to construct a
distributed observer for a continuous-time system  with a strongly
connected neighbor graph, which is capable of
 estimating state exponentially fast at a pre-assigned rate.
It is straightforward to modify this observer  to deal with
discrete-time systems.

\vspace{-0.1in}

An interesting idea, suggested in \cite{Kim2016CDC}, seeks to
simplify the structure of a distributed estimator for a
continuous-time system at the expense of some design flexibility.
This is done, in essence, by  exploiting the $A$-invariance of  the
unobservable spaces of the pairs $(C_i,A)$; this in turn enables one
to ``split'' the estimators into two parts, one based on
conventional spectrum assignment techniques and the other
 based on consensus  \cite{Kim2016CDC, trent, trent2,Lili19ACC,Lee2020}.
Reference \cite{Kim2016CDC} addresses the problem in continuous time
for undirected, connected neighbor graphs. The work of \cite{trent,
trent2} extends the result of \cite{Kim2016CDC} to the case when the
neighbor graph is directed and strongly connected. Establishing
correctness  requires one to choose gains to ensure   that certain
LMIs hold.
In \cite{Lee2020}, motivated by the distributed least squares solver problem, a modified algorithm which can deal with  measurement noise is proposed . 
In \cite{Lili19ACC} a simplified version  of the ideas in
\cite{trent} is presented. Because the ``high gain'' constructions
used in \cite{trent} and  \cite{Lili19ACC} don't apply
 in discrete-time, significant modifications are required to  exploit these ideas in a discrete-time context \cite{CDC19.1}.
 
\vspace{-0.1in}

Despite the preceding advances, until the appearance of
\cite{CDC17.1}, which first outlines the idea presented in this
paper,
  there were almost no  results
 for doing state estimation with time varying neighbor graphs for either discrete-time or
 continuous-time linear systems. For sure,
 there were a few partial results. For example,
 \cite{Acikmese2014}  suggests a  distributed observer using a  consensus filter
  for the state estimation  of discrete-time linear distributed system  for specially
  structured, undirected  neighbor graphs. Another example, in \cite{Ugrinovskii2013},
   an $H_\infty$ based observer is described  which is intended to function
    in the face of  a time-varying graph with
a Markovian randomly varying topology. 
It is also worth mentioning \cite{sanfelice}  which tackles  the
challenging problem of trying to define a distributed observer which
can function correctly in the
 face  of intermittent disruptions in available information. Although the problem addressed in \cite{sanfelice}
  is different
  than the problem to which this paper is addressed, resilience  in the face of intermittent disruptions
  is to some extent similar
   to the notion of resilience addressed in this paper.

\vspace{-0.1in}

The first paper to provide  a definitive solution to the distributed
state estimation problem for
 time varying neighbor graphs
 under reasonably relaxed assumptions was presented, in abbreviated form at the 2017 IEEE Conference on Decision and Control \cite{CDC17.1}.
The central contribution of \cite{CDC17.1} and this paper is  to
describe  a distributed observer for a  jointly observable,
 continuous-time linear  system
 with a time-varying neighbor graph $\mathbb{N}$ which is  capable  of estimating the system's state
 exponentially  fast at any prescribed rate. Assuming ``synchronous operation'', the only requirement
  on the graph is that it be strongly
 connected for all time.

\vspace{-0.1in}

Since the appearance of \cite{CDC17.1}, several other distributed
observers have been suggested which are capable of doing  state
estimation
 in the face of changing neighbor graphs. For example,  expanding on earlier work in \cite{MitraPurdue2016},
 \cite{Mitra2018a} provides a  procedure
 for constructing such an observer  which exploits in  some  detail the structure of $\mathbb{N}$  and its relation to
 the structure of the data matrices
  defining the system.  The resulting algorithm, which is tailored exclusively to discrete-time systems,
deals with state estimation under assumptions which are weaker than
strong connectivity. Recently we have learned that the split
spectrum observer idea first proposed in \cite{Kim2016CDC} and later
simplified in \cite{trent} and \cite{Lili19ACC} can be
 modified
 to deal with
strongly-connected time-varying neighbor graphs, although  only for
continuous time systems. See \cite{split2} for an unpublished report
on the subject.



\vspace{-0.1in}
\subsection{Organization}
\vspace{-0.1in}
The remainder of the paper is organized as follows.
The hybrid
observer itself is  described in
 \S\ref{sec:ob} subject to the assumption that all $m$ agents share the same event time sequence.
 Two cases are considered, one in which the interchanges of information  between agents
 are performed synchronously and the other case being when it is not. The synchronous case is the one  most comparable to
 the versions of the distributed observer problem treated  in \cite{Olfati-Saber2005} - \cite{Ugrinovskii2013}.
 The main result for this case is Theorem \ref{T1} which asserts that so long as the neighbor
  graph is strongly connected for all time, exponential convergence to zero at a prescribed convergence rate  of all  $m$
  state estimation errors is achieved. This is a new result which has no counterpart in any of the
  previously cited references.
    The same result is achieved in the asynchronous case \{cf. Theorem \ref{T2}\},
  but to reach this conclusion it is necessary to
   assume that the neighbor graph changes  in a suitably defined sense\footnote{It is worth noting at this point that many
   of the subtleties  of asynchronous operation are obscured  or at least  difficult to recognize
     in a discrete-time setting where there is
    invariably a single underlying  discrete-time clock shared by all $m$ agents.}
   These two theorems are the main contributions of this paper.
Their proofs can be found in \S\ref{analysis}.

  The aim of \S\ref{mismatch} is to explain what happens if  the assumption that all $m$ agents
  share the same event time sequence  is not made.
  For simplicity, this is only done for the case when
  differing event time sequences are the only cause of asynchronism.
   As will be seen, the consequence of event-time sequence mismatches
    turns out to be more of a robustness issue than an issue
  due to  unsynchronized  operation. In particular, it will become
 apparent that
if different agents use slightly different event time sequences then
asymptotically correct state estimates will not be possible unless
$A$ is a stability matrix, i.e., all the eigenvalues of matrix $A$ have strictly negative parts. While at first glance this may appear to
be a limitation of the distributed
 observer under consideration, it is in fact a limitation of virtually {\em all} state  estimators, distributed or not,
 which are not used in  feedback-loops. Since this easily established observation
  is apparently not widely appreciated, an explanation is  given at the end of the section.

By a (passively) {\it  resilient} algorithm for a distributed
process is meant an algorithm which, by exploiting built-in network
and data redundancies, is able to continue to function correctly in
the face of abrupt changes in the number of vertices and arcs in the
inter-agent communication graph upon which the algorithm depends. In
\S \ref{resilience}, it is  briefly explained how to configure
things  so that  that the proposed estimator can
 cope  with the situation when there is an arbitrary abrupt
change in the topology of the neighbor graph such as the loss or
addition of an arc or a vertex provided connectivity  is not lost in
an appropriately defined sense. Dealing with a loss or addition of
an arc proves to  to be easy to accomplish because of the ability of
the estimator to deal with time-varying graphs.  Dealing with the
loss or addition of a vertex
 is much more challenging and for this reason only preliminary results are presented.
Finally in \S \ref{Sec:simulation} simulation results are provided
to illustrate the observer's performance.

\vspace{-0.1in}

\section{Hybrid Observer}\label{sec:ob}
\vspace{-0.1in}

The overall hybrid observer to be considered consists of $m$
private estimators, one for each agent. Agent $i$'s private
estimator, whose function is to  generates an estimate $x_i$ of
$x$,  is  a hybrid dynamical system consisting of  a ``local
observer'' and a ``local parameter   estimator.'' The purpose of
local observer  $i$  is to generate an asymptotically correct
estimate of $L_ix$ where $L_i$ is any pre-specified,  full-rank
matrix
 whose kernel equals the kernel of the observability matrix of the pair $(C_i,A)$; roughly speaking,
 $L_ix$ can be thought of as that ``part of $x$'' which is observable to agent $i$.  Agent $i$'s
 {\em local observer}   is then an $n_i$-dimensional continuous-time, linear system of the form
\eq{ \dot{w}_i = (\bar{A}_i+ K_i\bar{C}_i)w_i-K_iy_i\label{po}}
where $n_i = \rank L_i$, $K_i$ is a gain matrix to be specified,
 and $\bar{C}_i$ and $\bar{A}_i$ are unique solutions
to the equations $C_i=\bar{C}_iL_i$ and $L_iA = \bar{A}_iL_i$,
respectively. As is well known, the pair $(\bar{C}_i,\bar{A}_i)$  is
observable and the {\em local observer  estimation error}
$\bar{e}_i \dfb w_i-L_ix$ satisfies
$$
\bar{e}_i(t) = e^{(\bar{A}_i+
K_i\bar{C}_i)t}\bar{e}_i(0),\;\;\;t\in[0,\infty)$$ Since
$(\bar{C}_i,\bar{A}_i)$ is an  observable pair, $K_i$ can be
selected so that $\bar{e}_i(t) $ converges to $0$
exponentially fast at any pre-assigned rate. We assume that each
$K_i$ is so chosen. Since \eq{w_i(t) = L_ix(t)
+\bar{e}_i(t),\;\;\;i\in\mathbf{m},\;\;\;t\in[0,\infty)\label{e1}
} $w_i$ can be viewed as a signal which approximates $L_ix$ in the
face of
 exponentially decaying additive noise, namely
$\bar{e}_i$.
 
 \vspace{-0.1in}
The other sub-system comprising agent $i$'s private estimator,
 is a ``local parameter estimator'' whose function is to generate estimates of $x$ at each of agent $i$'s preselected
 {\em event times} $ t_{i1}, t_{i2}, \ldots $. Here
$t_{i0}, t_{i1}, t_{i2}, \ldots $ is an ascending sequence of  event
times   with
 a fixed
spacing  of $T>0$ time units  between any two successive event
times. In this section it is assume that
 $t_{i0} = 0,\;i\in\mathbf{m}$, and consequently that all event time sequences are the
  same\footnote{It is easy to generalize the results  in this section to the case when
  event times are not evenly spaced provided that the spacings between successive pairs of
  event times  remains positive and  bounded.}. Thus
  $t_{is} = sT,\;s\geq 0,\;\;\;i\in\mathbf{m}.$
Between event times, each $x_i$ is generated using the equation
\begin{equation}
    \dot{x}_i=Ax_i \label{eq:1}
\end{equation}
Motivation for the development of the local parameter estimator
whose purpose is to enable agent $i$ to estimate $x(t_{i(s-1)})$
over
 the event time interval $[t_{i(s-1)},\;t_{is})$, stems from the fact that the equations
 \eq{w_j(t_{i(s-1)}) = L_jp +\bar{e}_j(t_{i(s-1)}),\;\;\;\;j\in\mathbf{m}\nonumber}
admit a unique solution, namely $p=x(t_{i(s-1)})$. Existence follows
from \rep{e1} whereas uniqueness
 is a consequence of the assumption of joint observability.

\begin{figure}[ht]
\centerline{\includegraphics [ width=0.4\textwidth]{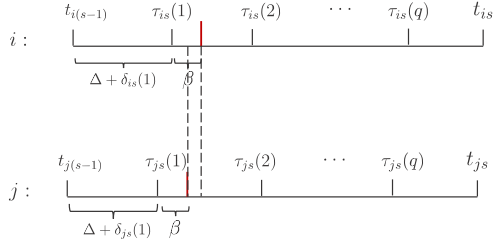}}
 \caption{A broadcast by
any agent $j$ at time $\tau_{js}(k-1) + \beta $ will occur within
the   {\em reception interval} $ [\tau_{is}(k-1), \;\;
\tau_{is}(k))$ of agent $i$.}
   \label{fig:timescale}
\end{figure}

The existence and uniqueness of $p$ suggest that an approximate
value of $x(t_{i(s-1)})$ can be obtained
  after a finite  number of iterations - say $q$ - using the linear equation solver discussed in\cite{LiliACC16}. Having obtained
   such an  approximate value of $x(t_{i(s-1)})$,  denoted below by $z_{is}(q)$, the desired estimate of $x(t_{is})$ can
    be taken as
\eq{   x_i(t_{is})\dfb e^{AT}z_{is}(q)\label{lll}} This is the
architecture which will be considered.
\\The computations needed to update   agent $i$'s  estimate of
$x(t_{i(s-1)})$  are  carried  out by agent $i$ during the event
time interval $[t_{i(s-1)}, t_{is})$. This is done using a local
parameter estimator which generates a sequence of  $q$ auxiliary
states $z_{is}(1),z_{is}(2),\ldots,z_{is}(q)$ where $q>0$ is a
positive integer to be specified below. The sequence is initialized
by setting \eq{z_{is}(0) =x_i(t_{i(s-1)}), \label{w1}} and  is
recursively  updated by agent $i$  at {\em local iteration   times}
$\tau_{is}(k),k\in\mathbf{q}\dfb\{1,2,\ldots, q\},$ known only to
agent $i$. It is assumed that the $\tau_{is}(k)$ together with the
initialization $ \tau_{is}(0)$  are of the form
\eq{\tau_{is}(k)=t_{i(s-1)} + k\Delta+
\delta_{is}(k),\;\;\;k\in\{0,1,\ldots,q\}\label{brp}} where
$\delta_{is}(0), \delta_{is}(1), \delta_{is}(2),\ldots,
\delta_{is}(q)$ is a sequence of small deviations
  which satisfy
\eq{\delta_{is}(k)  \in
[-\epsilon_i,\epsilon_i],\;\;\;k\in\{0,1,\ldots,q\},\label{dope}}
Here  $\epsilon_i$ is a small
   nonnegative number whose constraints will be described below and $\Delta$ is a positive number
    satisfying $$\Delta q+\max_i\{\epsilon_i,\;i\in\mathbf{m}\} \leq T$$
The signal
  $z_{is}(q)$ is agent $i$'s updated  estimate of
$x(t_{i(s-1)})$   and is  used to define $x_i(t_{is})$
as in \rep{lll}.
\\
The transfer of information between agents which is needed to
generate the $z_{is}(k)$, is carried out
 {\em asynchronously}
 as follows.
For $k\in\mathbf{q}$ and $j\in\mathbf{m}$, agent $j$ broadcasts
$z_{js}(k-1)$ at time
 $\tau_{js}(k-1) + \beta $
 where  $\beta $ is any prescribed nonnegative   number chosen  smaller than $\Delta $.
It is assumed
   that   the bounds  $\epsilon_i,\;i\in\mathbf{m}$,  appearing in \rep{dope}     are
 small enough   so that  there exist $\beta$ and $\Delta$ satisfying
\eq{\epsilon_i +\epsilon_j \leq\beta,\;\;\;\;\;\text{and}\;\;\;\;\;
\epsilon_i+\epsilon_j +\beta< \Delta,\;\;i,j\in\mathbf{m}
\label{top}}
These inequalities  ensure that for $k\in\mathbf{q}$, a broadcast by
any agent $j$ at time $\tau_{js}(k-1) + \beta $ will occur within
the   {\em reception interval} $ [\tau_{is}(k-1), \;\;
\tau_{is}(k))$ of agent $i$.
 Fig.~\ref{fig:timescale} provides an example of the update and communication times of two different different agents $i$ and $j$.
Accordingly, agent $j$ is a  data
source or just {\em  source }
 for agent $i$ on $[ \tau_{is}(k-1), \;\; \tau_{is}(k))$
if agent $j$ is in the reception range of agent $i$ at time
$\tau_{js}(k-1) +\beta $. Let $\mathcal{S}_{is}(k)$ denote the set of
labels of such agents; that is \eq{\mathcal{S}_{is}(k) =
\{j:j\in\mathcal{N}_i(\tau_{js}(k-1)+\beta)\}\label{source}} Note that
$i\in \mathcal{S}_{is}(k)$, for all $i\in\mathbf{m}$ so
$\mathcal{S}_{is}(k)$ is never empty. Clearly  agent $i$ can use the
signals $z_{js} (k-1)
  ,\;j\in\mathcal{S}_{is}(k)$, to compute $z_{is}(k)$.

\vspace{-0.1in}

Prompted by \cite{LiliACC16}, the update equation used to
recursively generate the $z_{is}(k)$ during agent $i$'s $s$th  event
time interval $[t_{i(s-1)},\;t_{is})$ is given by  
\begin{multline}  
      z_{is}(k)= \bar{z}_{is}(k-1) \\
 - Q_i(L_i\bar{z}_{is}(k-1)    - w_i(t_{i(s-1)})),\;\;\;\;k\in\mathbf{q}\;\; 
\label{kkk}\end{multline}
\noindent where 
   $Q_i = L_i'(L_iL_i')^{-1}$,  $\bar{z}_{is}(k-1)$ is an {\em averaged state}
\begin{equation}\label{eq:average}
\bar{z}_{is}(k-1 ) =
\frac{1}{m_{is}(k)}\sum_{j\in\mathcal{S}_{is}(k)}z_{js}(k-1),
\end{equation}
and
 $m_{is}(k)$
is the number of labels in $\mathcal{S}_{is}(k)$. The overall private
estimator for agent $i$ is thus described by the equations \rep{po}, \rep{eq:1}
- \rep{brp} and \rep{source} - \rep{eq:average}. 
  In summary, initialize $x_i(t_{i0})$, $w_i(0)$ randomly.
For $t\in [t_{i0},t_{i1})$, $\dot x_i=Ax_i$. Then for $s=1,2,\ldots$,
the algorithm of the hybrid estimator for anget $i$ is shown in Algorithm~\ref{alg:1}.
 
 \vspace{-0.1in}
 
\begin{algorithm}
  \caption{The hybrid estimator of agent $i$}\label{alg:1}
  \begin{algorithmic} [1]
      \State \textbf{Initialize } $x_i(t_{i0}),\;w_i(0),\; K_i,\;(\bar C_i,A_i),\;L_i,\;q$
      \State $\dot{w}_i = (\bar{A}_i+ K_i\bar{C}_i)w_i-K_iy_i$
       \For{$t\in [t_{i0},t_{i1})$}
      \State $ \dot x_i=Ax_i$ with  $x_i(t_{i0})$
      \EndFor
     \For{$s=1,2,\ldots$}
     \State $z_{is}(0) =x_i(t_{i(s-1)})$
      \For{ $k=1:q$}
        \State  Agent $i$ gets the sampled value $w_i(t_{i(s-1)})$ from its own estimator, and  receives $z_{js}(k-1)$ from its neighbor $j$.
       \[
   z_{is}(k)= \bar{z}_{is}(k-1)
-Q_i(L_i\bar{z}_{is}(k-1)    - w_i(t_{i(s-1)}))      
        \]
  where $\bar z_{is}$ is as defined in Eq.~\eqref{eq:average}
      \EndFor
      \For{$t\in [t_{is},t_{i(s+1)})$}
      \State $ \dot x_i=Ax_i$ with  $x_i(t_{is})=e^{AT} z_{is}(q)$
      \EndFor
     \EndFor
     \State {\bf Output:} $x_i$
  \end{algorithmic}
\end{algorithm}
In order to  complete the definition of the hybrid observer, it is
necessary to specify values of the
 $K_i$ and $q$. Towards this end, suppose that as a design goal it is desired to pick the $K_i$ and $q$ so that
 all $m$ {\em state estimation errors} \eq{e_i \dfb  x_i-x,\;i\in\mathbf{m}\label{see}} converge to zero
  as fast as $e^{-\lambda t}$ does
  where $\lambda > 0$
 is some  desired convergence rate.
 The $K_i$ would then have to be chosen using spectrum assignment or some other technique
 so that  the matrix exponentials $e^{(\bar{A}_i +K_i\bar{C}_i)t}$ all converge to
 zero at least as fast as  $e^{-\lambda t}$ does. This of course can be accomplished because each matrix
   pair $(\bar{C}_i,\bar{A}_i)$ is observable. In the sequel it will be assumed that for some preselected
    positive number $\bar{\lambda}>\lambda $, the $K_i$ have been chosen so that  for $i\in\mathbf{m}$ the local observer estimation errors satisfy
\eq{||\bar{e}_i(t)|| \leq
c_ie^{-\bar{\lambda}(t-\mu)}||\bar{e}_i(\mu)||,\;\;\;t\geq
\mu\geq 0 \label{sound}}
 where the  $c_i,\;i\in\mathbf{m}$ are nonnegative constants and
 $||\cdot||$ denotes the two-norm.

\vspace{-0.1in}

To  describe how to define an appropriate value of $q$ to attain the
desired convergence rate for the state estimation errors
$e_i,\;i\in\mathbf{m}$,   it is necessary to  take some preliminary
steps.   First, for each $i\in\mathbf{m}$, let $P_i$ denote the
orthogonal projection on the unobservable space of $(C_i,A_i)$. It
is easy to see that 
$P_i =I-
L_i'(L_iL_i')^{-1}L_i,\;\;i\in\mathbf{m}$. Moreover,  because of the
assumption of joint observability,
\eq{\bigcap_{i\in\mathbf{m}}\text{image}\; P_i =\{0\}\label{inter}} Next,
let $\mathcal{C}$ denote the set of all products
  of the form $P_{i_1}P_{i_2}\cdots P_{i_{(m-1)^2+1}}$ where each projection matrix in $\{P_i:i\in\mathbf{m}\}$
   occurs in each of such product  at least once.
    Note that $\mathcal{C}$ is a closed subset of $\R^{n\times n}$. 
 Since each projection matrix $P_i,\; i\in\mathbf{m}$ has a two-norm which is no greater than $1$, each matrix   $M\in \mathcal{C}$  has a two-norm less than or equal to $1$. Thus  $ \mathcal{C}$ is also a bounded and thus compact subset.
   In fact, each product in $\mathcal{C}$
actually  has two-norm strictly  less than $1$. This is a
consequence of  \rep{inter} and the requirement that each
 matrix in $\{P_i:i\in\mathbf{m}\}$ must occur in each product in $\mathcal{C}$ at least once \{Lemma 2,
 \cite{lineareqn}\}.
These observations imply that maximum  of the two-norms of the
matrices in $\mathcal{C}$, namely \eq{\rho \dfb
\max\{||M||:M\in\mathcal{C}\}, \label{woo}} exists  and is a real
non-negative number strictly  less than $1$ \footnote{It is worth
noting that although the matrices $L_i$ used in defining the $P_i$
are not uniquely determined by the unobservable spaces of the pairs
$(C_i,A)$, the
 orthogonal projection matrices  $P_i$  nonetheless are. Thus  the set  $\mathcal{C}$ used  in
 the definition of $\rho$ in  \rep{woo}
ultimately depends only on the family of unobservable spaces
of the pairs $(C_i,A),
 \;i\in\mathbf{m}$ and not on the particular manner in which the $L_i$ are chosen.
 Just how to explicitly characterize this dependence is a topic for future research.}. This
in turn implies that the {\em attenuation constant} \eq{\alpha\dfb
1- \frac{(m-1)(1-\rho)}{m^{(m-1)^2}}\label{atten}}
is also a real  non-negative number strictly  less than $1$. 
As will  become evident below \{cf. \rep{phi} and Lemma
\ref{cold}\}, in the {\em idealized case} when
 all $\epsilon_i$ and $\bar{e}_i$
are zero,
 for any integer $p>0$ and any given  value of $q$ satisfying
\eq{q \geq p((m-1)^2+1), \label{sww}}
 the value of the signal
$$ \max\{||z_{is}(k)-x(t_{i(s-1)})||:  i\in\mathbf{m}\}$$
is attenuated by at least a factor $\alpha^{p}$ after $q$
iterations during each  event - time interval
$[t_{i(s-1)},\;t_{is})$; i.e.,  for $s\geq 1$,
\begin{eqnarray*}\lefteqn{\max\{||z_{is}(q)-x(t_{i(s-1)})||:  i\in\mathbf{m}\}\leq}\;\;\;\;\;\;\;\;\;\\
 & & \alpha^{p}\max\{||z_{is}(0)-x(t_{i(s-1)})||:  i\in\mathbf{m}\}\end{eqnarray*}
It will soon be apparent, if it is not already from \rep{lll},
\rep{w1} and \rep{see}, that over each  event- time interval
$[t_{i(s-1)},\;t_{is})$,
\begin{eqnarray}\lefteqn{\max\{||e_i(t_{is})||:  i\in\mathbf{m}\}\leq}\;\;\;\;\;\;\;\;\;\nonumber\\
 & & e^{||A||T}\alpha^{p}\max\{||e_i(t_{i(s-1)})||:  i\in\mathbf{m}\}\label{koop}\end{eqnarray}
Since each event - time   interval is of length $T$, to achieve an
 exponential convergence rate of $\lambda$  in the idealized case,
 it  is necessary to pick $q$ so that
\rep{sww} holds where $p$  is any integer  satisfying $
e^{||A||T}\alpha^{p}< e^{-\lambda T}$. In other words, the
requirement on $q$ is that \rep{sww} hold where \eq{p > \left
\lceil\frac{(\lambda +||A||)T}{\ln(\frac{1}{\alpha})}\right
\rceil,\label{bill}} with $\lceil r\rceil$ here denoting,
 for any nonnegative number $r$,
  the smallest integer $k\geq r$.
The following theorem, which applies to the {\em synchronous case}
when all of the $\epsilon_i$ are zero, \{but not necessarily   the
$\bar{e}_i$\}
 summarizes these observations.
\begin{theorem}{\bf Synchronous case:}
 Suppose $\epsilon_i = 0,\;\;i\in\mathbf{m}$, and that the neighbor
 graph $\mathbb{N}(t)$ is strongly connected for all $t$. Let $\rho $ and $\alpha $ be defined by \rep{woo} and
 \rep{atten} respectively.  Then each state estimation $e_i= x_i-x,\;i\in\mathbf{m}$,  of the hybrid observer
defined by \rep{po}, \rep{eq:1}
- \rep{brp} and \rep{source} - \rep{eq:average},
tends to zero as fast as $e^{-\lambda t}$ does provided $q$
satisfies \eq{q> ((m-1)^2+1)\left \lceil \frac{(\lambda +||A||)
T}{\ln(\frac{1}{\alpha})}\right \rceil\label{carol}}
\label{T1}\end{theorem}
\vspace{-0.2in}
\noindent This theorem will be proved in the next section.
Several comments are in order. First, the attenuation  of
$\max\{||e_i||:i\in\mathbf{m}\}$  by $\alpha ^{p}$ over an event
time interval  is not likely
 to be tight and a larger attenuation constant can almost certainly be expected. This is important because the
larger the attenuation constant the smaller the required value of
$q$ needed to achieve a given  convergence rate. Second, the
hypothesis that $\mathbb{N}(t)$ strongly connected is almost
certainly stronger than is necessary, the notion of a repeatedly
jointly  strongly connected sequence of graphs \cite{lineareqn}
being a likely less stringent alternative.

\vspace{-0.1in}

To deal with the asynchronous case when at least some of the
$\epsilon_i$ are nonzero,
 it is necessary to assume that
  $\mathbb{N}(t)$ is  constant on  each of the time intervals
\begin{multline}\mathcal{I}_{s}(k) =[-\epsilon +
sT +(k-1)\Delta + \beta, \epsilon + sT \\+(k-1)\Delta + \beta ],
\;k\in\mathbf{q},\;s\geq 1\label{disjoint}\end{multline} where
\eq{\epsilon = \max\{\epsilon_i:i\in\mathbf{m}\}\label{aaa}} For
this assumption to make sense, these intervals cannot overlap.
 The following lemma   establishes that this is in fact the case.

\begin{lemma}\label{snow} Suppose that $q\geq 2$ and that
the $\epsilon_i$ are fixed  nonnegative numbers satisfying  the
constraints in \rep{top}.
  Then for each $s\geq 0$,
 the $q$ time  intervals defined by \rep{disjoint}
are non-overlapping and each is a subinterval  of $ [sT, (s+1)T)$.
 \end{lemma}

\vspace{-0.1in}

\noindent{\bf Proof of Lemma \ref{snow}:} Fix $2\leq k \leq q$. Note
that $2\epsilon <\beta$ because of \rep{top}. This implies that
$\epsilon +sT +(k-2)\Delta +\beta <-\epsilon +sT +(k-1)\Delta+\beta$
and thus that $\mathcal{I}_{s}(k)$ and $\mathcal{I}_{s}(k-1)$ are disjoint.
Since this holds for all $k$ satisfying $2\leq k \leq q$, all
$\mathcal{I}_{s}(k),\;k\in\mathbf{q}$ are disjoint.

\vspace{-0.1in}

From \rep{top}, $\epsilon \leq\beta$ and $\epsilon +\beta <\Delta$.
These inequalities imply that $-\epsilon +sT +\beta \geq sT$ and
$\epsilon +sT +(q-1)\Delta +\beta <(s+1)T$ respectively. From this
it follows that $\mathcal{I}_{s}(1)\subset [sT, (s+1)T)$, that
$\mathcal{I}_{s}(q)\subset [sT, (s+1)T)$ and thus that
$\mathcal{I}_{s}(k)\subset [sT, (s+1)T),\;\;k\in\mathbf{q}$. $\qed$

\vspace{-0.1in}


We are led to the asynchronous version of Theorem \ref{T1}.

\begin{theorem}{\bf Asynchronous case:}
 Suppose the  $\epsilon_i,\;i\in\mathbf{m}$, satisfy \rep{top} and  that the neighbor
 graph  $\mathbb{N}(t)$ is 
constant
 on  each interval
$\mathcal{I}_{s}(k),\;k\in\mathbf{q},\;s\geq 1$ and strongly connected for all $t$. Let $\rho $ and $\alpha $ be defined by \rep{woo} and
 \rep{atten} respectively and suppose that $q$ satisfies \rep{carol}.  Then as in Theorem \ref{T1},
  each state estimation $e_i= x_i-x,\;i\in\mathbf{m}$,
 of the hybrid observer
defined by\rep{po}, \rep{eq:1}
- \rep{brp} and \rep{source} - \rep{eq:average},
 tends to zero as fast as $e^{-\lambda t}$ does.
\label{T2}\end{theorem}
\vspace{-0.1in}
The proof of this theorem will be given in the next section.
 Notice that the asynchronous case here can not be recognized in a discrete-time setting with a discrete-time  clock shared by all $m$ agents considering delays\cite{RR2}.
\vspace{-.1in}
 \subsection{\bf A special case}\label{spec}
 \vspace{-.1in}
 It is possible to relax somewhat  the lower bound \rep{carol} for $q$ to
achieve exponential convergence
  in the  special case when the neighbor graph $\mathbb{N}(t)$ is symmetric and strongly
   connected for all $t$. 
This can be accomplished by replacing the straight averaging rule
defined by
 \rep{eq:average},  with the convex combination rule
\eq{\bar{z}_{is}(k-1)=(1-\frac{m_{is}(k)}{m+1})z_{is}(k-1) +\frac{1}{m+1}
 \!\!\!\!\sum_{j\in \mathcal{S}_{is}(k)} \!\!\!\!z_{js}(k-1)\label{amon}}
where $m_{is}(k)$ is the number of labels in $\mathcal{S}_{is}(k)$.

To proceed, let $\mathcal{G}$ denote the set of all symmetric and strongly connected,
 graphs on $m$ vertices.
Each graph
$\mathbb{G}\in\mathcal{G}$ uniquely determines a  matrix
$M_{\mathbb{G}} =I - \frac{1}{m+1}L_{\mathbb{G}}$ where $L_{\mathbb{G}}$ is the
 Laplacian of the  simple, weakly connected  \{undirected\} graph determined by $\mathbb{G}$.
 It is easy to see that
 $M_{\mathbb{G}}$ is a symmetric, doubly stochastic matrix
 with positive diagonals and that $\mathbb{G}$ is its graph.
 The connection between these  matrices and the  update rule defined by \rep{amon} will become apparent later
 when assumptions are made which enable us to identify the subsets $\mathcal{S}_{is}(k)$  appearing in \rep{amon} with the neighbor sets
of the neighbor graph $\mathbb{N}(((s-1)T+(k-1)\delta +\beta)$ \{c.f. Lemmas \ref{arm} and \ref{farm}\}.
Later in this paper it will also be shown that
 $P(M_{\mathbb{G}}\otimes I)$  is a
  contraction in the two norm \{Lemma \ref{grass}\}.
This means that \begin{equation} \label{eq:specialsigma}\sigma
=\max_{\mathbb{G}\in\mathcal{G}}||P(M_{\mathbb{G}}\otimes
I)||\end{equation}
is a nonnegative number less than  one.

As will become clear,   to achieve a convergence rate of $\lambda$,
 it is sufficient  to pick $q$  large enough to that
$e^{||A||T}\sigma^q<e^{-\lambda T}$.  In other words, in the special case when $\mathbb{N}(t)$ is
symmetric and strongly connected for all time,
 instead
of choosing $q$   to satisfy \eqref{carol}, to achieve an exponential
convergence rate of $\lambda$  it is enough to choose
 $q$  to satisfy the less demanding constraint
  \begin{equation}\label{eq:q2}q > \left\lceil  \left ( \frac{(\lambda
+||A||)T}{\ln(\frac{1}{\sigma})}\right )\right \rceil
\end{equation}
Justification for this claim is  given in
\S\ref{analysis}.
Choosing $q$ in this way   is easier that choosing $q$ according to \rep{carol} because the computation of $\sigma $
 is less demanding than the computation of $\rho $ and consequently $\alpha$. On the other hand,
 this special approach only applies when the neighbor graph is symmetric.


\vspace{-.1in}

\section{Analysis}\label{analysis}

\vspace{-.1in}

The aim of this section is to analyze the behavior of the hybrid
observer defined in the last section. To do this,
 use will be made of the notion of a ``mixed matrix norm'' which will now be defined.
 For any  positive integers $k$, $m$, $n$ and $p$, let $\mathcal{M}$  denote the real $kmnp$- dimensional
   vector space  of block  partitioned matrices
    $\mathbf{M} =[M_{ij}]_{k\times m}$ where  each block $M_{ij}$ is a $n\times p $ matrix.
     By the {\em mixed matrix norm}
    of $\mathbf{M}\in\mathcal{M}, $ written $|\mathbf{M}|$, is meant the infinity norm of the matrix
     $[||M_{ij}||]_{k\times m}$ where $||M_{ij}||$
     is the two-norm of $M_{ij}$. For example, with $e$ denoting
     the ``stacked'' state estimation error
     $e\dfb \text{column}\{e_1,e_2,\ldots,e_m\}$ the quantity  $\max\{||e_i||:i\in\mathbf{m}\}$ mentioned in
     the last section,
      is $|e|$,  the mixed matrix
      norm  of $e$.
       It is straight forward to verify that $|\cdot |$ is in fact a norm
        and  that
       this norm is sub-multiplicative \cite{lineareqn}.



       Recall that  the  purpose of agent $i$'s local parameter
 estimator   defined by \rep{w1}, \rep{kkk}, and \rep{eq:average} is to estimate $x(t_{i(s-1)})$
 after executing  $q$ iterations
  during  the   $s$th event time interval of agent $i$. In view of this,
   we   define the {\em parameter error vectors} for $i\in\mathbf{m}$,
\eq{\pi_{is}(k) = z_{is}(k) - x(t_{i(s-1)}) \;\;\;k = 0,1,\ldots,
q\label{pe}} for  all $s\geq 1$. This, \rep{w1},
and the definition of $e_i$ in \rep{see}  imply that \eq{\pi_{is}(0)
=  e_i(t_{i(s-1)}),\;\;s\geq 1,\;\;i\in\mathbf{m}\label{sn1}} In
addition,  from  \rep{eq:1}, \rep{lll} and   \rep{see} it is clear
that that \eq{e_i(t_{is}) = e^{AT}\pi_{is}(q), \;s\geq
1,\;\;i\in\mathbf{m}\label{sn2}}
\vspace{-.1in}

To derive the update equation for $\pi_{is}(k)$ as $k$
 ranges from $1$ to $q$,  we first  note  from  \rep{eq:average} that
\begin{multline}\bar{z}_{is}(k-1) -x(t_{i(s-1)}) =
\frac{1}{m_{is}(k)}\!\!\!\!\sum_{j\in\mathcal{S}_{is}(k)}\!\!\!\!\pi_{js}(k-1)\;\label{painn}\end{multline}
Next note that  because of \rep{e1} and \rep{kkk}
\begin{multline*}\pi_{is}(k) = \bar{z}_{is}(k-1) -x(t_{i(s-1)})\\
-Q_i(L_i\bar{z}_{is}(k-1)-x(t_{i(s-1)})
-\bar{e}_i(t_{i(s-1)}))\end{multline*} From this and
\rep{painn} it follows that for $k\in\mathbf{q}, \;i\in\mathbf{m},\;s\geq 1$,
\begin{multline}\pi_{is}(k)\!\! = \!\!\frac{1}{m_{i(s)}(k)}P_i\!\!\!\!\!\sum_{j\in\mathcal{S}_{is}(k)}\!\!\!\!\!\pi_{js}(k-1)
 +Q_i\bar{e}_i(t_{i(s-1)})\label{sn3}\end{multline}
where as before, $P_i =I- Q_iL_i$.
 These are the {\em local parameter error equations} for the hybrid observer.


The next step in the analysis of the system is studied the evolution
of the {\em all-agent parameter
 error vector}
 $$\pi_s(k) = \text{column}\{\pi_{1s}(k), \pi_{2s}(k),\ldots, \pi_{ms}(k)\}$$
Note first that because of \rep{sn1} and \rep{sn2} \eq{\pi_s(0) =
e(s-1),\;\;s\geq 1\label{jane}} and \eq{e(s) =
e^{\tilde{A}T}\pi_s(q),\;s\geq 1\label{shift}} where $e(s)$ is the
 {\em all-agent  state estimation error vector}
\eq{e(s)= \text{column}\{e_1(t_{is}),\ldots,
e_m(t_{ms})\},\;\;s\geq 0\label{allagent}} and
 $\tilde{A} = \text{diagonal}\{A,A,\ldots, A\}$.

In order to develop an update equation for
  $\pi_{s}(k)$ as $k$ ranges from $1$ to $q$, it is necessary
to combine the
  $m$ update equations in \rep{sn3} into a single  equation and to do this requires a succinct
  description of the graph determined by the sets
  $\mathcal{S}_{is},\;i\in\mathbf{m}$ defined in \rep{source}. There are two cases to consider:
    the synchronous case which is when all of the $\epsilon _i = 0$
     and the asynchronous case when some or all of the $\epsilon_i$ may be non-zero.
     The following lemmas cover both cases.

\begin{lemma} {\bf Synchronous Case: } Suppose $\epsilon_i = 0,\;i\in \mathbf{m}$. Then for
 any fixed value
 of $\beta$ satisfying \rep{top}, including $\beta =0$,
 \begin{equation}\mathcal{S}_{is}(k) \!=\! \mathcal{N}_i((\!s\!-\!1\!)T  \!+\!(k-1)\Delta\! +\! \beta),
i\in\mathbf{m},
  k\in\mathbf{q}\label{rat}\end{equation}\label{arm}\end{lemma}
 \vspace{-0.2in}
\noindent{\bf Proof of Lemma \ref{arm}:} By hypothesis all
$\epsilon_i = 0$.
 Clearly \rep{top} can be satisfied with $\beta = 0$. Moreover from  \rep{brp} and \rep{dope} and the assumption that
 $t_{i(s-1)} = (s-1)T$, it follows that
  $\tau_{js}(k-1) = (s-1)T +(k-1)\Delta,\;j\in\mathbf{m}$, so
 $\mathcal{N}_i(\tau_{js}(k-1) + \beta)  =
       \mathcal{N}_i((s-1)T  +(k-1)\Delta + \beta)$. From this and \rep{source} it follows that
       \rep{rat} is true. $\qed $


 The following lemma asserts that \rep{rat} still holds in the  asynchronous case when some of the $\epsilon_i$ are nonzero,
 provided  $\mathbb{N}(t)$ is 
constant
 on  each interval
$\mathcal{I}_{s}(k),\;k\in\mathbf{q},\;s\geq 1$.

\begin{lemma} If the $\epsilon_i$ satisfy the constraints in  \rep{top} and
 $\mathbb{N}(t)$ is
 constant
 on  each interval
$\mathcal{I}_{s}(k),\;k\in\mathbf{q},\;s\geq 1$ then \rep{rat} is true.
\label{farm}\end{lemma}
\noindent{\bf Proof of Lemma \ref{farm}:} Fix $i\in\mathbf{m}$,
$s\geq 1$ and $k\in\mathbf{q}$.
In light of \rep{brp}, \rep{dope} and the assumption that
$t_{i(s-1)} = (s-1)T$, it is clear that for any  $j\in\mathbf{m}$,
 \begin{multline*}\tau_{js}(k-1) +\beta \in
 [-\epsilon_j +(s-1)T + (k-1)\Delta +\beta,\\
  \epsilon_j +(s-1)T+ (k-1)\Delta +\beta]\subset\mathcal{I}_{(s-1)}(k)\end{multline*}
  Moreover, $(s-1)T+(k-1)\Delta +\beta \in \mathcal{I}_{(s-1)}(k)$. But by assumption,
  $\mathbb{N}(t)$ is constant on $\mathcal{I}_{(s-1)}(k)$ which means that $
  \mathcal{N}_i(t)$ is constant
  on $\mathcal{I}_{(s-1)}(k)$.Therefore  $\mathcal{N}_i( \tau_{js}(k-1) +\beta)
   =\mathcal{N}_i((s-1)T+ (k-1)\Delta +\beta)$. From this and the definition of $\mathcal{S}_{is}(k)$
   in \rep{source}, it follows that \rep{rat} is true. $\qed$

In summary,    Lemmas \ref{arm} and \ref{farm} assert that \rep{rat}
holds in the synchronous case when all $\epsilon_i = 0$, or
alternatively in the asynchronous case when the neighbor graph
$\mathbb{N}(t)$ is  constant
 on  each interval
$\mathcal{I}_{s}(k),\;k\in\mathbf{q},\;s\geq 1.$ Because  of this, the following
steps to obtain an update equation for $e(s)$ apply to both cases.

Equation \rep{rat}  implies that the  graphs determined by the
$\mathcal{S}_{is}(k),\; \;k\in\mathbf{q}, \;i\in\mathbf{m},\;s\geq 1$
are the neighbor graphs
 $\mathbb{N}((s-1)T +(k-1)\Delta +\beta ),\;k\in\mathbf{q}, \;s\geq 1$. Since $\mathbb{N}(t)$ is assumed to be strongly connected for
  all $t\geq 0$, each of these neighbor graphs is strongly connected. These graphs are used as follows.

  Let $\mathcal{G}$ denote the set of all directed
 graphs on $m$ vertices which have self-arcs at all vertices.
 Note that $\mathcal{G}$ is a finite set and that $\mathbb{N}(t)\in\mathcal{G},\;t\geq 0$. Each
 graph $\mathbb{G}\in\mathcal{G}$ uniquely determines a so-called ``flocking-matrix'' which is an
 $m\times m$ stochastic matrix of the form
 $D^{-1}_{\mathbb{G}}A'_{\mathbb{G}}$, where  $A_{\mathbb{G}}$ and
  $D_{\mathbb{G}}$ are respectively the adjacency matrix
 and   diagonal in-degree matrix of of $\mathbb{G}$; $D_{\mathbb{G}}$ is nonsingular because
 each graph in $\mathcal{G}$ has self-arcs at all vertices.

For $k\in\mathbf{q}$ and $s\geq 1$, let $F_s(k)$ denote  the
flocking matrix  determined by
 $\mathbb{N}((s-1)T +(k-1)\Delta +\beta)$.
 Then \rep{sn3} implies  for $s\geq 1$  that
\eq{\pi_s(k) = P(F_s(k)\otimes I)\pi_s(k-1)+
  Q\bar{e}(s-1),\;\;k\in\mathbf{q}\label{nu1}}
 where \eq{\bar{e}(s) =\text{column}\{\bar{e}_{1}(t_{1s}),\ldots,
  \bar{e}_m(t_{ms})\},\;\;s\geq 0\label{funch}}
$P=\text{diagonal}\{P_1,\ldots,P_m\},
 Q=\text{diagonal}\{Q_1,\ldots,Q_m\}$, 
 and $I$ is the $n\times n$ identity.
Thus for $s\geq 1$, \eq{\pi_s(q) = \Phi_s(0)\pi_s(0) +\left
(\sum_{k=1}^q\Phi_s(k)\right )Q\bar{e}(s-1)\label{rain}}
where $\Phi_s(k)$ is the state transition matrix defined by
\eq{\Phi_s(k) =  P(F_s(q)\otimes I)\cdots P(F_s(k+1)\otimes
I)\label{phi}} for $0\leq k<q$ and by $\Phi_s(q) = I$ for $k=q$.
From this, \rep{jane} and \rep{shift} it follows that for $s\geq 1$,
the all-agent state estimation error $e(s)$ satisfies \eq{e(s) =
A(s)e(s-1) +B(s)\bar{e}(s-1),\;\;s\geq 1\label{ef}} where
\begin{eqnarray}
A(s) &=& e^{\tilde{A}T}\Phi_s(0),\;\;s\geq 1\label{a}\\
B(s) &=& e^{\tilde{A}T}\sum_{k=1}^q\Phi_s(k)Q,\;\;s\geq 1\label{b}
\end{eqnarray}
To determine the convergence properties
 of   $e(s)$   as $s\rightarrow \infty$ use will be made of the following
  lemma which gives bounds on the norms of the coefficient matrices $A(s)$ and $B(s)$
  appearing in \rep{ef}.

\begin{lemma}
Suppose that $q$ satisfies the inequality given in Theorem \ref{T1}.
Then
\begin{eqnarray}
|A(s)| &\leq &e^{-\lambda T},\;\;s\geq 1 \label{dep}\\
|B(s)| &\leq &  qe^{||A||T}|Q|,\;\;s\geq 1\label{dep2}
\end{eqnarray}\label{crunch}\end{lemma}
\vspace{-0.35in}
In order to justify the bound on the norm of
    $A(s)$ given in \rep{dep}, use will be made of the following   lemma.
 which is a simple variation on a   result in \cite{lineareqn}.

\begin{lemma}Let  $\mathcal{F}$ denote the set of all
  flocking matrices determined by those
   graphs in $\mathcal{G}$ which are strongly connected.
 For any set of $\mu\geq(m-1)^2 +1 $ flocking matrices
$S_1,S_2,\ldots , S_{\mu}$  in $\mathcal{F}$ \eq{|P(S_{\mu}\otimes
I)P(S_{\mu-1}\otimes I) \cdots P(S_1\otimes I)|\leq
\alpha\label{jack}} where  $\alpha $ is the attenuation constant
\vspace{-0.1in}
$$\alpha= 1-\frac{(m-1)(1-\rho)}
{m^{(m-1)^2}}$$ \label{cold}\end{lemma}
\vspace{-0.3in}
 \noindent{\bf Proof of Lemma \ref{cold}:} Fix  $\mu\geq(m-1)^2+1$, set $k = (m-1)^2$ and let
and $S_1,S_2,\ldots , S_{\mu}$ be flocking matrices in $\mathcal{F}$.
    Then
    \vspace{-0.in}
\begin{multline*}P(S_{\mu}\otimes I)P(S_{\mu-1}\otimes I) \cdots P(S_2\otimes I)P(S_1\otimes I)\\
=\{P(S_{\mu}\otimes I)\cdots P(S_{k+2}\otimes
I)\}\\\{P(S_{k+1}\otimes I) \cdots
 P(S_2\otimes I)P\}\{S_1\otimes I\}\end{multline*}
But for any flocking matrix $S\in\mathcal{F}$, $|S\otimes I|
=||S||_{\infty}=1$ where $||\cdot ||_{\infty}$ is the infinity norm.
From this, the sub-multiplicative property of the mixed matrix norm,
and the fact that $|P|\leq 1$, it follows that
 \vspace{-0.1in}
\begin{multline*}|P(S_{\mu}\otimes I)P(S_{\mu-1}\otimes I) \cdots
P(S_2\otimes I)P(S_1\otimes I)|\\
\leq|P(S_{k+1}\otimes I) \cdots
 P(S_2\otimes I)P|\end{multline*}
 \vspace{-0.1in}
 In view of equation (26) of \cite{lineareqn},
$$|P(S_{k+1}\otimes I) \cdots P(S_2\otimes I)P|\leq 1-\frac{(m-1)(1-\rho)}
{m^{(m-1)^2}}$$ \vspace{-0.1in}Therefore \rep{jack} is true. $\qed $

\noindent{\bf Proof of Lemma \ref{crunch}:} Lemma \ref{cold}
implies that if for a given  integer $p>0$, if
 $q \geq p((m-1)^2+1)$ then for any
$s\geq 1$, \eq{|(P(F_{s}(q)\otimes I)\cdots
P(F_{s}(1)\otimes I)|\leq \alpha^p\label{jack2}} Therefore by
\rep{phi} and \rep{a}, if $q$ is so chosen, then $|A(s)|\leq
e^{||A||T}\alpha^p $. Thus by picking $p$ so large that \eq{
e^{||A||T}\alpha^p <e^{-\lambda T}\label{fst}} and then setting  $q=
p((m-1)^2+1) $ one gets \rep{dep}.
 The requirement on $p$ determined  by \rep{fst} is equivalent to
 the requirement on $p$ determined  by \rep{bill}. It follows that \rep{dep} will hold provided $q$ satisfies
   the inequality given in Theorem \ref{T1}.
\vspace{-0.2in}

Recall that $|P|\leq 1$ and that  $|S\otimes I| = 1$ for any
$m\times m$ stochastic matrix $S$. From this and the
sub-multiplicative property of the mixed matrix norm it follows that
the matrix $\Phi_s(k)$ defined by \rep{phi} satisfies
\eq{|\Phi_s(k)|\leq 1,\;\;\;k\in\mathbf{q},\;\;s\geq 1\label{mail}}

This and the
 definition of $B(s)$  in \rep{b}  imply that for all $s\geq  1$,
$|B(s)|\leq qe^{||A||T}|Q|$. Thus \rep{dep2} is true.
 $\qed $

It is obvious  at this point that because \rep{rat} holds in both the
synchronous and asynchronous cases,
  the same arguments can be used to prove both Theorem \ref{T1} and Theorem \ref{T2}.

\vspace{-0.1in}
\noindent{\bf Proof of Theorems \ref{T1} and \ref{T2}:} In view of
\rep{ef} and Lemma \ref{crunch} it is possible to write
$$|e(s)| < e^{-\lambda T}|e(s-1)| + b |\bar{e}(s-1)|,\;s\geq 1
$$
where  $b =qe^{||A||T}|Q|$. Therefore \eq{|e(s)| < e^{-\lambda
sT}|e(0)|
 +  b  \sum _{k=1}^se^{-\lambda (s-k)T}|\bar{e}(k-1)| \label{bing}}
\\
To deal with the term involving $\bar{e}$ in \rep{bing}, we
proceed as follows. Note first  from \rep{sound} that
$$||\bar{e}_i(t_{is})||\leq  c_ie^{-\bar{\lambda }T}||\bar{e}_i(t_{i(s-1)})||,\;\;i\in\mathbf{m},\;\;\;s\geq 1$$
Thus
$||\bar{e}_i(t_{is})||\leq  c_ie^{-\bar{\lambda }sT}
||\bar{e}_i(t_{i0})||$ for $i\in\mathbf{m}\;s\geq 1$. It
follows from this and the definition of $\bar{e}(s)$  in
\rep{funch} that
$$|\bar{e}(s)| \leq  ce^{-\bar{\lambda }sT}|\bar{e}(0)|,\;\;s\geq1$$
where $c=\max\{c_i,\;i\in\mathbf{m}\}$. Thus for $s\geq 1$
\begin{multline*}b\sum_{k=1}^s e^{-\lambda (s-k)T}|\bar{e}(k-1)|\\\leq
       cb\sum _{k=1}^se^{-\lambda (s-k)T}e^{-\bar{\lambda}(k-1)T}|\bar{e}(0)| \\
= cb e^{-(\lambda
s-\bar{\lambda})T}\sum_{k=1}^se^{-(\bar{\lambda}-\lambda)kT)}|\bar{e}(0)|
\\
\leq cb e^{-(\lambda s-\bar{\lambda})T}\sum_{k=1}^{\infty}e^{-(\bar{\lambda}-\lambda)kT)}|\bar{e}(0)|\\
=cb e^{-(\lambda s-\bar{\lambda})T}\frac{e^{-(\bar{\lambda}-\lambda)T}}{1-e^{-(\bar{\lambda}-\lambda)T}}|\bar{e}(0)|\\
= cb e^{-\lambda sT}\frac{e^{\lambda
T}}{1-e^{-(\bar{\lambda}-\lambda)T}}
|\bar{e}(0)|\hspace{.83in}\end{multline*} Using \rep{bing}
there follows \eq{|e(s)|\leq e^{-\lambda sT}(|e(0)|+
d|\bar{e}(0)| ),\;s\geq 1  \label{flum}} where
$$d = cb \frac{e^{\lambda T}}{1-e^{-(\bar{\lambda}-\lambda)T}}$$

Fix $i\in\mathbf{m}$. In view of \rep{flum} and the definition of
$e(s)$ in \rep{allagent},
$$||e_i(t_{i(s-1)})||\leq e^{-\lambda (s-1)T}(|e(0)|+ d|\bar{e}(0)| )\;\;i\in\mathbf{m},\;\;s\geq 1$$
But for $t\in(t_{i(s-1)},\;t_{is})$,  $\dot{x}_i = Ax_i$;
consequently $\dot{e}_i = Ae_i$ for the same values of $t$.
Therefore
$$e_i(t) = e^{A(t-(s-1)T)}e_i(t_{i(s-1)}),\;\;t\in[t_{i(s-1)},\;t_{is}),\;\;s\geq 1$$
so
$$||e_i(t)||\leq e^{||A||T}||e_i(t_{i(s-1)}||,\;\;t\in[t_{i(s-1)},\;t_{is}),\;\;s\geq 1$$
Therefore  for $t\in[t_{i(s-1)},t_{is})$ and $s\geq 1$
$$||e_i(t)|| \leq e^{(||A||T-\lambda (s-1)T)}(|e(0)|+ d|\bar{e}(0)| )$$
Now for $i\in\mathbf{m}$,
$$e^{-\lambda sT} \leq  e^{-\lambda  t},  \;\;t\in[t_{i(s-1)},\;t_{is})$$
so
$$||e_i(t)||\leq e^{(||A||T-     \lambda  t)}(|e(0)|+ d|\bar{e}(0)| ),  \;\;t\in[(t_{i(s-1)},\;t_{is})$$
Since this holds for all $s\geq 1$
$$||e_i(t)||\leq e^{(||A||T    -  \lambda t)}(|e(0)|+ d|\bar{e}(0)| ),\;\;t\geq 0$$
which proves that the state estimation errors
$e_i,\;i\in\mathbf{m}$, all converge to zero as fast as $e^{-\lambda
t}$ does. $\qed $

\subsection{Special case}

We now turn to the special case mentioned in \S\ref{spec}. In this
case the definition of the state-transition matrix $\Phi$ appearing
in \rep{rain} changes from \rep{phi} to
\begin{equation}\label{eq:newphi}
    \Phi_s(k)=P(W_s(q)\otimes I)\cdots P(W_s(k+1)\otimes
I)
\end{equation}
for $0\leq k<q$, and $W_s(k) \df M_{\mathbb
N((s-1)T+(k-1)\Delta+\beta)}$ with graph   $\mathbb
N((s-1)T+(k-1)\Delta+\beta)$.

 Although the formula for $e(s)$, namely
 \rep{ef}, and the definitions of  $A(s)$ and $B(s)$  in \rep{a}
and \rep{b} are as before, the bounds for $A(s)$ and  and $B(s)$
given by \rep{dep} and \rep{dep2} no longer apply. To proceed, use will be made of the following
lemma.

\begin{lemma} Let $F$ be an $m\times m$ doubly stochastic matrix with positive diagonals and a strongly connected graph.
 Suppose that $P_i,\;\;i\in\mathbf{m}$, is a set of $n\times n$
 orthogonal projection matrices such that
 \vspace{-0.1in}
\eq{\bigcap_{i=1}^m\text{image}\;P_i =0\label{po2}}
Then the matrix $P(F\otimes I)$ is a contraction in the $2$-norm where
 $P = \text{\rm diag\;}\{P_2,P_2,\ldots,P_m\}$.
\label{grass}\end{lemma}
 \vspace{-0.1in}

\noindent{\bf Proof:} Write $S$ for $F\otimes I$ and note that $S$ is doubly stochastic with positive
 diagonals and a strongly connected graph.
Since $||P||\leq 1$, it must be true that
  that
 $||PS||\leq ||S|| $. Moreover $||S||\leq 1$ because $S'S$
 is stochastic; thus $||PS||\leq 1$. Hence it is enough to prove that $||PS||\neq 1$ or equivalently that
 $||S'P|| \neq 1$
\\
Suppose that $||S'P|| =1$ or equivalently that
   $PSS'Px = x$ for some nonzero vector $x$. Clearly PSS'Px = Px which implies that  $Px = x$ and thus that
   $x'SS'x = x'x.$  Therefore $||S'x|| = ||x||$.  From this and Lemma 1 of \cite{ACC16.1}   it follows that
   $SS'x=x$.
 Now $SS'$ is  stochastic.  Moreover its graph is strongly connected because
 $S$ has a strongly connected graph and positive diagonals, as does $S'$. Thus by the Perron Frobenius theorem,
 $SS'$ has exactly one eigenvalue at  $1$ and all the rest must be inside the unit circle; in addition the
  eigenspace for the eigenvalue $1$ must be spanned by the one-vector $\mathbf{1}_{nm}$. Therefore $x=\mu\mathbf{1}_{nm}$
  for some nonzero scalar $\mu$. Therefore $P\mathbf{1}_{nm} = \mathbf{1}_{nm}$ which implies that
  $\mathbf{1}_n = P_i\mathbf{1}_n,\;\;i\in\mathbf{m}$, But this is impossible because of \rep{po}. $\qed$

\vspace{-0.1in}

 The following
lemma gives the bounds on $A(s)$ and $B(s)$ for the special case
under consideration.

\begin{lemma}\label{lemma:special}
Suppose that $q$ satisfies  \rep{eq:q2}. Then
\begin{eqnarray} &\ & \|A(s)\|\leq e^{-\lambda T},\;\; s\geq 1\label{eq:newA}\\
&\ &\|B(s)\|\leq qe^{\|A\|T}\|Q\|,\;\; s\geq
1\label{eq:newB}\end{eqnarray}
\end{lemma}
\vspace{-0.1in}
\noindent{\bf Proof:} Lemma \ref{grass} implies that for each $0\leq
k<q$, $\|P(W_s(k)\otimes I)\|< 1$. Moreover, $\|P(W_s(k)\otimes
I)\|\leq \sigma$ where $\sigma$ is chosen according to
\eqref{eq:specialsigma}. From this and the sub-multiplicative
property of the two norm it follows that
 \[
\|P(W_s(q)\otimes I)\cdots P(W_s(1)\otimes I)\| \leq \sigma^{q}\]
Therefore by \rep{a} and \eqref{eq:newphi}, if $q$ is so chosen to
satisfy \rep{eq:q2}, then
\[
\|A(s)\|\leq e^{\|A\|T}\sigma ^{q}< e^{-\lambda T}
\]
Thus \eqref{eq:newA} is true. Recall that $\|P(W_s(k)\otimes I)\|<
1$ and  $\|W_s(k)\otimes I\|\leq 1$ for $0\leq k<q$. From this and
the sub-multiplicative property of the two norm, the matrix
$\Phi_s(k)$ defined by \eqref{eq:newphi} satisfies
\[\|\Phi_s(k)\|\leq 1, \;\; k\in \mathbf{q},\;\; s\geq 1\]
This and \rep{b} imply \eqref{eq:newB}. \qed

Other than the modifications in the bounds on $A(s)$ and $ B(s)$
given in  the above lemma, everything else is the same for both the
synchronous and asynchronous versions of the problem. So what one
gains  in this special case is exponential convergence at a
prescribed rate with a smaller value of $q$.

\vspace{-0.1in}

\section{Event-time Mismatch - A Robustness Issue}\label{mismatch}
\vspace{-0.1in}

In the preceding section it was shown that the hybrid observer under
 discussion will function correctly if local iterations are
 performed synchronously across the network no matter how fast the associated neighbor graph changes,
  just so long as it is always strongly connected.
  Correct performance is also assured in the face of asynchronously
  executed local  iterations across the network during each event time interval, provided the neighbor graph changes
    in a suitably defined sense. Implicitly assumed in these two cases is
   that the  event time sequences  of all $m$ agents are the same.
  The aim of this section is to explain what happens if this assumption is not made.
  For simplicity, this will only be done for the case when
  differing event time sequences are the only cause of asynchronism.
   As will be seen, the consequence of event-time sequence mismatches
    turns out to be more of a robustness issue than an issue
  due to  unsynchronized  operation. In particular, it will become
 apparent that
if different agents use slightly different event time sequences then
asymptotically correct state estimates will not be possible unless
$A$ is a stability matrix. While at first glance this may appear to
be a limitation of the distributed
 observer under consideration, it is in fact a limitation of virtually {\em all} state  estimators, distributed or not,
 which are not used in  feedback-loops. Since this easily explained observation
  is apparently not widely appreciated, an explanation of this simple fact will be given at the end of this section.

\vspace{-0.1in}

There are two differences between  the setup to be considered here
and the setup considered  in the last section. First it will now  be
assumed that the local deviation times $\delta _{is}(k)$ appearing
in \rep{brp} are all zero. Thus in place of \rep{brp} the local
iteration times for agent $i$  on $[t_{i(s-1)},t_{is})$
\eq{\tau_{is}(k) = t_{i(s-1)}+k\Delta, \;k\in\{0,1,\ldots,
m\}\label{brbn}} Second instead of assuming  that the
initializations $t_{i0}$  of the $m$ agents' event time sequences
are all zero, it will be assumed instead that each $t_{i0}$ is a
small number known only to agent $i$ which lies in the interval
 $[-\epsilon_i,\epsilon_i]$ where, as before,  $\epsilon_i$ is a small nonnegative number.
 This means that even though the
 event time sequences of all $m$ agents are
 still periodic with period $T$,  the sequences are not synchronized with each other.
As before it is assumed that within  event time interval
$[t_{j(s-1)},t_{js})$,
 agent  $j$ broadcasts iterate $z_{js}(k-1)$ at time $\tau_{js}(k-1) +\beta$.
 To ensure that this time falls within  the reception interval $[t_{i(s-1)},t_{is})$ of each agent $i$, it will continue to
  be assumed  that \rep{top}  holds.
Apart from these modifications the setup to be considered here  is
the same as the one considered previously. As a consequence, many of
the steps in the analysis of the hybrid observers performance
 are the same  as they were for the previously considered case.

\vspace{-0.1in}

Our first objective is to develop  the relevant equations for the
local parameter error vector
 $\pi_{is}(k)$ defined by \rep{pe}.  Although \rep{sn1} and \rep{sn2} continue to hold without change, \rep{sn3}
 requires modification. To  understand what needs to be changed, it is necessary  to first derive a relationship between
$x(t_{i(s-1)})$ and $x(t_{j(s-1)})$.  Towards this end, note that
$$x(t_{i(s-1)})\!=\!x(t_{j(s-1)}) +x((s-1)T +t_{i0})-x((s-1)T +t_{j0}) $$
because $t_{k(s-1)} = t_{k0} +(s-1)T$ for all $k\in\mathbf{m}$. From
this and \rep{eq:1} it follows that
$$x(t_{i(s-1)}) =x(t_{j(s-1)}) +\left (e^{At_{i0}}-e^{At_{j0}}\right )x((s-1)T)$$
Hence  \rep{eq:average}  can now be used to obtain
\begin{multline}\bar{z}_{is}(k-1) -x(t_{i(s-1)}) =
\frac{1}{m_{is}(k)}\sum_{j\in\mathcal{S}_{is}(k)}\pi_{j(s-1)}(k-1)\\
+\Gamma_{is}(k) x((s-1)T)\;\;\;\;\;\;\;\label{painn2}\end{multline}
where \eq{\Gamma_{is}(k) =
\frac{1}{m_{is}(k)}\sum_{j\in\mathcal{S}_{is}(k)} \left
(e^{At_{i0}}-e^{At_{j0}}\right )\label{favmeet}} Next note that
because of \rep{e1} and \rep{kkk}
\begin{multline*}\pi_{is}(k) = \bar{z}_{is}(k-1) -x(t_{i(s-1)})\\
-Q_i(L_i\bar{z}_{is}(k-1)-x(t_{i(s-1)})
-\bar{e}_i(t_{i(s-1)}))\end{multline*} From this and
\rep{painn2} it follows that
\begin{multline}\pi_{is}(k) = \frac{1}{m_{i(s)}(k)}P_i\sum_{j\in\mathcal{S}_{is}(k)}\pi_{js}(k-1)
 +Q_i\bar{e}_i(t_{i(s-1)})\\+P_i\Gamma_{is}(k)x((s-1)T),\;\;k\in\mathbf{q}, \;i\in\mathbf{m},\;s\geq
 1\label{sn3n}\end{multline}
which is the modified version of \rep{sn3} needed to proceed.  The
difference between \rep{sn3} and  \rep{sn3n}
 is  thus the inclusion in \rep{sn3n} of the  term
$P_i\Gamma_{is}(k)x((s-1)T)$.

\vspace{-0.1in} 
The assumption that the event time sequences of the agents  may
start at a different time requires  us to make the same assumption
as before about the neighbor graph $\mathbb{N}(t)$,
 namely that it is  constant
 on  each interval
$\mathcal{I}_{s}(k),\;k\in\mathbf{q},\;s\geq 1$. The assumption makes sense in the present context  for the same reason as before, specifically
 because the interval
 $\mathcal{I}(k)$ defined by \rep{disjoint} do not overlap. This, in turn,  is because the bounds  $\epsilon_i$
 have been assumed to satisfy
  \rep{top} which guarantees that Lemma \ref{snow} continues to hold.

The next step in the analysis of the hybrid observer is to   study
the evolution of the all-agent parameter
 error vector
 $$\pi_s(k) = \text{column}\{\pi_{1s}(k), \pi_{2s}(k),\ldots, \pi_{ms}(k)\}$$
As before, \rep{jane} and \rep{shift} continue to hold where $e(s)$
is the all - agent state estimation error defined by \rep{allagent}.
A simple modification of the proof of
 Lemma \ref{farm} can be used to establish the lemma's validity  in the present context. Consequently
  a proof will not be given.
The lemma enables us  to combine the individual update equations
 in \rep{sn3n}, thereby obtaining  the update equation
\begin{multline*}\pi_s(k) = P(F_s(k)\otimes I)\pi_s(k-1)+
  Q\bar{e}(s-1)\\ +P\Gamma_s(k)x((s-1)T),\;\;k\in\mathbf{q}\end{multline*}
 where
 \eq{\Gamma_s(k) = \text{column}\{\Gamma_{1s}(k),\ldots,
 \Gamma_{ms}(k)\}\label{roman}}
The steps involved in doing this are essentially the same as the
steps involved in deriving  \rep{nu1}. Not surprisingly,
 the only difference between \rep{nu1} and  \rep{roman}  is the inclusion in the latter  of the
 term $P\Gamma_s(k)x((s-1)T)$.

From  \rep{jane},  \rep{shift}, and \rep{roman}  it follows at once
that the all-agent state estimation error vector satisfies
\begin{multline}e(s) = A(s)e(s-1) +B(s)\bar{e}(s-1) \\+ G(s)x((s-1)T),\;\;s\geq 1\label{efn}\end{multline}
where $A(s)$ and $B(s)$ are  as defined in \rep{a} and \rep{b}
respectively,  and
$$G(s) = e^{\tilde{A}T}\sum_{k=1}^q\Phi_s(k)P\Gamma_s(k)$$
The following lemma gives a bound on the mixed matrix norm of
 $G(s)$.

\begin{lemma}
Suppose that $q$ satisfies the inequality given in Theorem \ref{T1}.
Then \eq{|G(s)| \leq 2mq\epsilon
||A||e^{||A||(T+\beta)}\label{dep3}}\label{crunch2}\end{lemma}
Note that this bound is small when $\epsilon $ is small. This means
that small deviations of the agent's event time sequences from the
nominal event time sequence $0, T, 2T,\ldots $ produce small
effects on the
 error dynamics in \rep{efn}, provided of course $x$ is well behaved; i.e., $A$ is a stability matrix!
 More will be said about this point below.

\noindent{\bf Proof of Lemma \ref{crunch2}:} From  \rep{favmeet},
$$||\Gamma_{is}(k)||\leq \sum_{j\in\mathbf{m}} || e^{At_{i0}}-e^{At_{j0}} ||$$
In general, for any real square matrix $M$, and real numbers $t$ and
$\tau$
$$||e^{Mt} - e^{M\tau}||\leq ||M(t-\tau)||e^{||Mt||}$$
so
$$||\Gamma_{is}(k)\|\leq \sum_{j\in\mathbf{m}} ||A(t_{i0}-t_{j0})||e^{||At_{i0}||}$$
By assumption $|t_{i0}|\leq \epsilon_i $ and $|t_{i0}-t_{j0}|\leq
\epsilon_i +\epsilon_j$. But $\epsilon_i\leq \beta$ and  $\epsilon =
\max\{\epsilon_i,\;i\in\mathbf{m}\}$. Thus $|t_{i0}|\leq \beta$ and
$|t_{i0}-t_{j0}|\leq 2\epsilon$. Therefore
$$||\Gamma_{is}(k)||\leq 2m\epsilon ||A||e^{||A||\beta}$$
so
$$|\Gamma_{s}(k)|\leq 2m\epsilon ||A||e^{||A||\beta}$$
In view of \rep{mail} and the definition of $G(s)$,  $|G(s)|\leq
qe^{||A||T}||\Gamma_s(k)||$. It follows that
 \rep{dep3} holds. $\qed $

Taking the construction leading to \rep{bing} as a guide, it is not
difficult to derive from \rep{efn} the inequality
\begin{multline}|e(s)|\leq e^{-\lambda sT}(|e(0)|+ d|\bar{e}(0)| )\\ +
\epsilon g \sum_{k=1}^se^{-\lambda(s-k)T}||x((k-1)T)||,\;s\geq 1
\label{flum2}\end{multline} where $d$ is as defined just below
\rep{flum} and $g =2mq ||A||e^{||A||(T+\beta)}$. Comparing
\rep{flum} to \rep{flum2}, we see that the effect of the change in
assumptions leads to the inclusion in \rep{flum2} of the term
involving $x$.

 At this point there are two distinct cases to consider -
 either $e^{At}$ converges to zero or it does not. Consider first the case when $e^{At}$ converges to zero.
Then there must be positive constants $c_a$ and $\lambda_a$ such
that $||e^{At}||\leq c_ae^{-\lambda_at}$.  By treating the term
involving  $x$ in  \rep{flum2} in the same manner as the term
involving $\bar{e}$ in \rep{flum} was treated, one can easily
conclude that
 for a suitably defined constant
$h$
$$|e(s)|\leq e^{-\lambda sT}(|e(0)|+ d|\bar{e}(0)| + \epsilon h ||x(0)|| ),\;\;s\geq 1 $$
if $\lambda _a >\lambda$, or
$$|e(s)|\leq e^{-\lambda sT}(|e(0)|+ d|\bar{e}(0)|) + \epsilon he^{-\lambda_a sT} ||x(0)||,\;\;s\geq 1  $$
if $\lambda_a\leq \lambda$. If  the former is true, then the same
arguments as  were used in the last section
  can be used to show that the
state estimations errors $e_i(t)$  converge to zero as fast as
$e^{-\lambda t}$ does. On the other hand, if the latter is true, by
similar reasoning $e_i(t)$  can easily be shown to
 converge to zero as fast as $e^{-\lambda_a t}$ does. Note that in this case, if  $\lambda_a$ is  small,
  the effect of the resulting slow convergence
  of $x$ will  to some extent be mitigated  by the smallness of  $\epsilon $, so even with
   small $\lambda_a$, the performance of the hybrid observer may be acceptable for sufficiently small
    perturbations of the start times of the event time sequences from $0$.
\vspace{-0.1in}

In the  other situation, which is when  $A$ is not a stability
matrix, the hybrid observer cannot perform acceptably except
possibly if finite time state estimation is all that is desired
 and $\epsilon $ is sufficiently small.

\vspace{-0.1in}
 \noindent{\bf Key Point:}   {\em This  limitation applies not only to the hybrid observer discussed
  in this paper,
 but to {\bf all } state estimators, centralized or not, including Kalman filters
 which are not being used in feedback
  loops.}\footnote{Some of the adaptive observers developed in the past may be an exception to this, but such observers
 invariably  require persistent  excitation to achieve exponential convergence.}

\vspace{-0.1in}
  Experience has shown that this  limitation is not widely recognized, despite its
  simple justification. Here is the justification.
\\
Suppose one is trying to obtain an  estimate   $\widehat{x}$  of the
state $x$
   of a single-channel,  observable linear system $y = Cx$, $\dot{x} = Ax$  using an observer
   but approximately  correct values of $A$ and $C$
     - say $\widehat{A}$ and $\widehat{C}$  - upon which to base the observer design are known.
     The observer would then be a linear system
     of the form
\eq{\dot{\widehat{x}}= \widehat{A}\widehat{x}
+K(\widehat{C}\widehat{x} - y )\label{ppp}}
     with $K$ chosen to exponentially stabilize $\widehat{A} +K\widehat{C}$.
Then it is  easy to see that  the state estimation error $e =
\widehat{x}-x$ must satisfy
    $$\dot{e} = (\widehat{A}+K\widehat{C})e + (\widehat{A}-A +K(\widehat{C} - C))x $$
Therefore if $A$ is not a stability matrix and either  $\widehat{A}$
is not exactly equal to $A$ or $\widehat{C}$ is not exactly equal
to  $C$,
 then instead of converging to zero,
 the state estimation error $e$
will   grow without bound
  for almost any initialization.
 In other words, with robustness in mind,  {\em the problem of trying to obtain an estimate of the state of
 a linear system with an ``open-loop'' state estimator,
 does not make sense unless $A$ is a stability matrix.}\;
Of course, if one is trying to use  a state estimator  generate an
estimate  $\widehat{x}$ of  the state $x$ of
 the forced linear system
$$\dot{x} = Ax +BF\widehat{x}$$  where $A+BF$ is a stability matrix, this problem does not arise, but to accomplish this
one has to change the estimator  dynamics defined in \rep{ppp}  to
$$\dot{\widehat{x}}= \widehat{A}\widehat{x} +K(\widehat{C}\widehat{x} - y ) +BF\widehat{x}$$
While this modification works in the centralized case, it cannot be
used in the decentralized case as explained in \cite{disfeed}. 
  In fact,  until recently there appeared to be only one  of
    distributed observer which could be used in a
   feedback configuration thereby
   avoiding the robustness issue just mentioned \cite{disfeed}. However, recent research suggests other approach may emerge \cite{Kim2021}.

\vspace{-0.1in}

\section{Resilience}\label{resilience}
\vspace{-0.1in}

By a (passively) {\it  resilient} algorithm for a distributed
process is meant an algorithm which, by exploiting built-in network
and data redundancies, is able to continue to function correctly in
the face of abrupt changes in the number of vertices and arcs in the
inter-agent communication graph upon which the algorithm depends. In
this section,it will be shown that the proposed estimator can
 cope  with the situation when there is an arbitrary abrupt
change in the topology of the neighbor graph such as the loss or
addition of an arc or a vertex provided connectivity  is not lost in
an appropriately defined sense.

\vspace{-0.1in}

Consider first the situation when there is a potential loss or
addition of $a$ arcs in the neighbor graph.
 Assume the neighbor graph is  {\em $\bar{a}$-arc
redundantly strongly connected} in that  the graph is strongly
connected and remains strongly connected after any $a\leq \bar{a}$
arcs are removed. With this assumption, strong connectivity of the
neighbor graph and
 jointly observability of the system are ensured when any $a\leq \bar{a}$
arcs  are lost. Alternatively, if any number of new arcs are  added,
strong connectivity and  joint observability  are clearly still
ensured. Thus, in the light of  Theorem \ref{T1}, whenever $a\leq
\bar{a}$ arcs are lost from
 or  added to the neighbor
graph, the hybrid estimator  under consideration will still function
correctly without
 the need for  any ``active'' intervention  such as redesign
  of any of the $K_i$ or readjustment of $q$. In fact, Theorem \ref{T1} guarantees
  that  correct performance will prevail, even if arcs  change over and over, no matter
  how fast, just so long as strong connectivity is maintained for all time.

\vspace{-0.1in}
 Consider next the far more challenging situation when  at some time $t^*$ there is a  loss  of $v<m$ vertices
  from the neighbor graph $\mathbb{N}(t).$ For this situation, only preliminary results currently exist. One possible
   way to deal with this situation  is as follows.

\vspace{-0.1in}
As a first step, pick the $K_i$ as before, so that all $m$ local
observer state estimator errors
 converge to zero as fast as $e^{-\lambda t}$ does.
 Next, assume  that the neighbor graph  is {\it $\bar{v}<m$-vertex
redundantly strongly connected} in that  it is strongly connected
and remains strongly connected after any $v \leq\bar{v}$ vertices
are removed. Assume in addition that the system described by
\rep{sys1}, \rep{sys2}
 is {\it
$\bar{v}$ redundantly jointly observable} in that
 the system  which results  after any $v\leq \bar{v}$ output measurements $y_i$ have been deleted,
 is still jointly observable. Let $\mathcal{D}$ denote the family of all nonempty subsets
 $\mathbf{d}\subset \mathbf{m}$ such that each subset $\mathbf{d}\in\mathcal{D}$ contains at
 least
   $m-\bar{v}$ vertices. Thus each loss of at most $\bar{v}$ vertices results in a strongly connected
   subgraph of $\mathbb{N}(t)$ for some subset $\mathbf{d}\in\mathcal{D}$;
   call this subgraph $\mathbb{N}_{\mathbf{d}}(t)$. Correspondingly, let
 $\Sigma_{\mathbf{d}}$ denote the multi-channel linear system which results when those outputs $y_i$
 $i\not\in\mathbf{d}$ are deleted from \rep{sys1}, \rep{sys2}. Thus  $\Sigma_{\mathbf{d}}$ is
  a jointly observable  multi-channel linear system whose channel outputs are the $y_i,\;i\in\mathbf{d}$. Fix $\lambda >0$.

\vspace{-0.1in}

Fix  $\mathbf{d} \in\mathcal{D}$ and let $m_{\mathbf{d}}$ denote the
number of vertices in $\mathbb{N}_{\mathbf{d}}$.
 Since $\Sigma_{\mathbf{d}}$ is jointly observable it is possible to compute a number
 $\rho_{\mathbf{d}}$ which satisfies \rep{woo}. Using the pair $(\rho_{\mathbf{d}}, m_{\mathbf{d}})$ in place of the pair
 $(\rho, m)$ in \rep{atten} and \rep{carol}, it is possible to calculate a value of $q$,
   for which \rep{carol} holds. In other words, for this value of $q$, henceforth labelled
   $q_{\mathbf{d}}$
  Theorem \ref{T1} holds for the multichannel system $\Sigma_{\mathbf{d}}$ and neighbor graph $\mathbb{N}_{\mathbf{d}}(t)$.
  By then picking $$q^* = \max_{\mathbf{d}\in\mathcal{D}}q_{\mathbf{d}} $$  one obtains a value of $q$ for which Theorem
  \ref{T1} holds for all pairs $(\Sigma_{\mathbf{d}},\mathbb{N}_{\mathbf{d}}(t))$ as $\mathbf{d}$ ranges over $\mathcal{D}$.
Suppose a hybrid observer using $q=q^*$  is implemented. Suppose in
addition  that at some time $t^*$, for some specific
$\mathbf{d}\in\mathcal{D}$,    agents with labels in
$\mathbf{m}-\mathbf{d}$ stop functioning. Clearly the remaining
agents with labels in $\mathbf{d}$ will be able to deliver the
desired
 state estimates with the prescribed convergence rate bounds.
 In this sense, the observer under consideration  is resilient to
   vertex losses. However, unlike the loss or addition of edges mentioned above,
   no claim is being made at this point  about what might happen if  some or all of the lost vertices
    rejoin the network, especially if this loss-gain process is  rapidly  reoccurring over and over as  time evolves.
\vspace{-0.1in}

 A similar approach can be used to deal with the situation when at
some time $t^*$, the network gains some additional agents. In this
case one would have to specify all possibilities and make sure that
for each one, one has a strongly connected graph and a jointly
observable system.

\vspace{-0.1in}

A little thought reveals that what makes it possible to deal with a
change in the number of vertices in this way, is the fact that there
is a single  {\em scalar} quantity, namely $q$,   with the property
that for each possible
  graphical configuration resulting from an anticipated  gain or loss of vertices, there is a value of $q$  large enough
  for the distributed observer to perform correctly and moreover if $q$ is assigned   the maximum of these values
  then the distribute  observer will perform correctly no matter which of the anticipated
  vertex changes is actually encountered.
 Since the distributed observers described in
 \cite{Kim2016CDC,trent,Lili19ACC}  also  require the  adjustment of only a single scalar-valued
 quantity for a given   neighbor graphs,
 the same basic idea just described   can be used to make the observers  in \cite{Kim2016CDC,trent,Lili19ACC}
resilient to  a one-time  gain or loss of the number of vertices
  on their associated neighbor graph.
  On the other hand, some distributed observers
  such as the ones described in
 \cite{martins,tac.17,R2} are not   really amenable to this kind of
generalization because  for such observers changes in network
topology  require completely new designs involving the change of
many of the observer's  parameters.
 There are also papers \cite{R1,R5} deal with sensor attacks, where a malicious attacker can manipulate their observations arbitrarily when each sensor only has one dimensional measurement.

\vspace{-0.1in}

\section{Simulation}\label{Sec:simulation}

\vspace{-0.1in}

\begin{figure}[h]
  \centering
  \begin{subfigure}[b]{0.23\textwidth}
 \centering
 \includegraphics [height=0.8in]{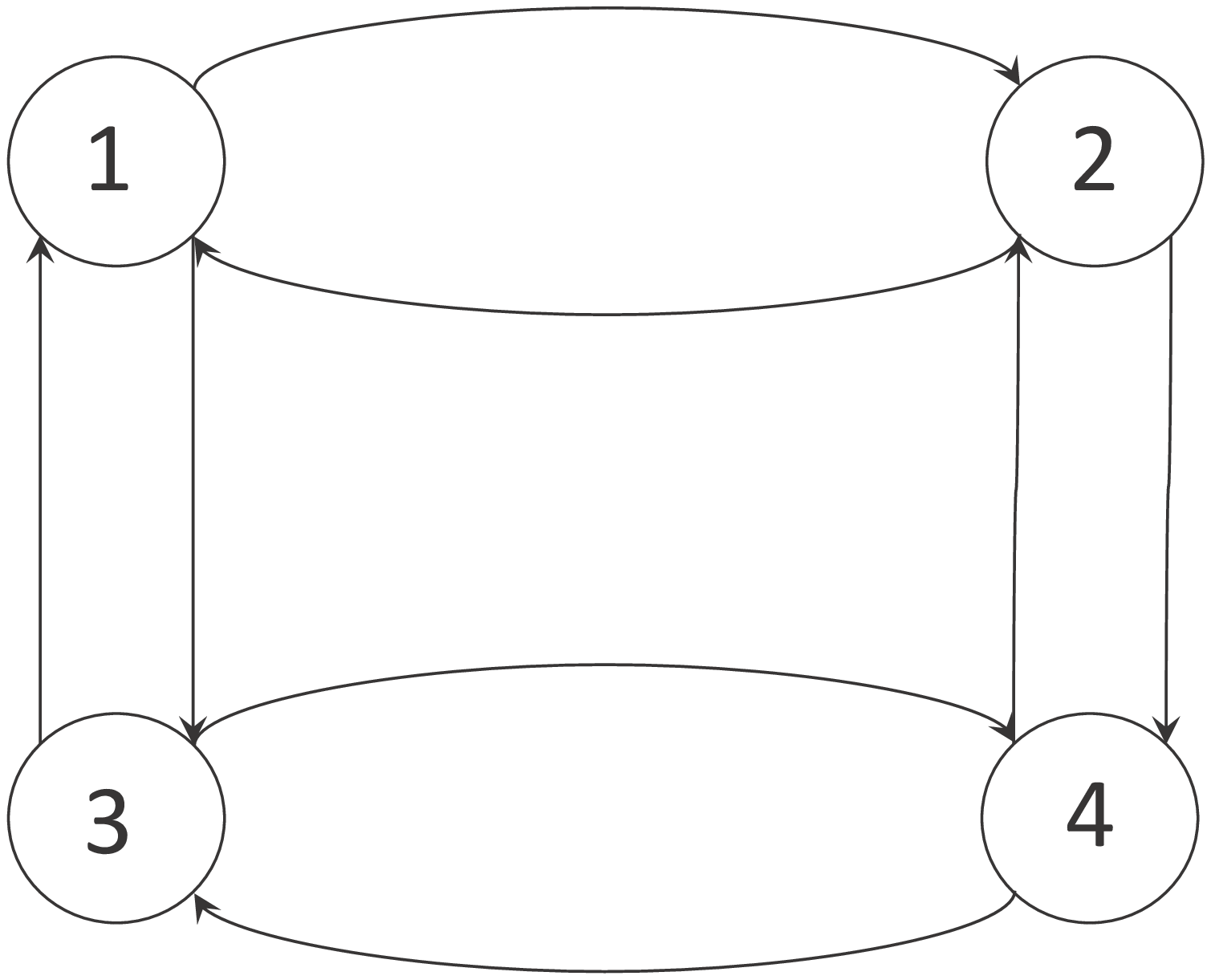}
   \caption{\ } 
   \label{lunch}
 \end{subfigure}
 \hfill
 \begin{subfigure}[b]{0.23\textwidth}
 \centering 
 \includegraphics [ height=0.8in]{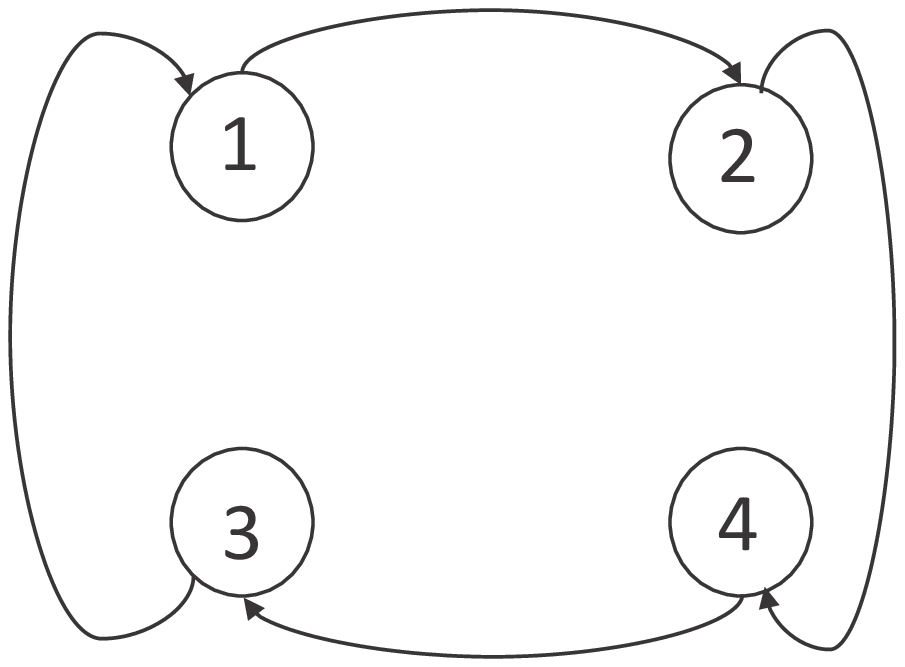}
   \caption{\ }
    \label{lunch2}
  \end{subfigure} 
  \caption{$\mathbb{N}(t)$}
\end{figure}

The following simulations are intended to illustrate (i) the
performance of the hybrid observer in the face of system noise, (ii)
the robustness of the hybrid observer with respect to variations of
event time  sequences,
 and (iii) resilience of the hybrid observer to the loss
 or   gain of an agent.
Consider the four channel, four-dimensional, continuous-time system
described by the equations  $\dot{x}=Ax,\; y_i=C_ix,\; i\in
\{1,\;2,\;3,\;4\}$, where
\vspace{-0.1in}
{\small $$A=\begin{bmatrix} -0.1& 0.4& 0 & 0\\ -0.1&-0.1&0&0\\ 0 &0 & -0.2 & 0.2\\
0 & 0 & -2 & 0.1\end{bmatrix} $$ }and $C_i$ is the $i$th unit row
vector in $\R^{1\times 4}$. Note that $A$ is a stable matrix with
two eigenvalues at $-0.1\pm j0.2$ and  a pair of complex eigenvalues
at $-0.05\pm j0.6144$.
 While the system is jointly observable, no single pair $(C_i,A)$  is observable. However the system
 is ``redundantly jointly observable'' in that  what remains after the
  removal of any one output $y_i$, is still jointly observable.
For the first two simulations $\mathbb{N}(t)$ is switching back and
forth between  Figure  \ref{lunch} and Figure~\ref{lunch2}, and for
the third simulation the neighbor graph is as shown in
Figure~\ref{lunch}. Both graphs are strongly connected, and the
graph in Figure~\ref{lunch} is redundantly strongly connected in
that it is strongly connected and
 remains strongly connected after any one vertex is removed.
Suppose $T=1$ for this system. To achieve a convergence rate of
$\lambda =2$, $\bar \lambda$  and $q$ are chosen to be  $q=50$ and
$\bar \lambda =3$  respectively. 
\\
\vspace{-0.1in}
\noindent For agent 1: $\bar{C}_1 = [0\; 1]$, 
  \[
  \bar{A}_1\!\!=\!\!\begin{bmatrix}-0.1 & -0.1\\ 0.4 & -0.1\end{bmatrix},\; L_1 \!\!= \!\!\begin{bmatrix} 0 & 1 & 0 & 0\\ 1 & 0 & 0 & 0
\end{bmatrix},\; K_1\!\! =\!\!
-\begin{bmatrix}13.7\cr 4.8
\end{bmatrix}\]
\vspace{-0.1in}
\\
\noindent For
agent 2:  $ \bar{C}_2= [0\; 1]$,
\vspace{-0.1in}
\[  \bar{A}_2=\begin{bmatrix}-0.1 & -0.4\\ 0.1 & -0.1\end{bmatrix}, L_2 =\begin{bmatrix}-1 & 0 & 0 & 0\\ 0 & 1 & 0 & 0 \end{bmatrix},  
K_2 =-\begin{bmatrix}54.7\\ 4.8\end{bmatrix}\]
\vspace{-0.1in}
\noindent
For agent 3:  $\bar{C}_3= [0\; 1]$,
\vspace{-0.1in}
\[\bar{A}_3=\begin{bmatrix}0.1 & -2 \\ 0.2 & -0.2\end{bmatrix},
 L_3= \begin{bmatrix}0 & 0 &0 &1\\ 0 & 0 & 1& 0\end{bmatrix}, K_3=-\begin{bmatrix}30.6\\
4.9\end{bmatrix}\]
\vspace{-0.1in}
  \noindent For agent 4:
$\bar{C}_4= [0\; 1]$,
\vspace{-0.05in}
\[\bar{A}_4=\begin{bmatrix}-0.2 & -0.2\\ 2 & 0.1\end{bmatrix},  L_4= \begin{bmatrix} 0 & 0 & -1 &0\\ 0 & 0 & 0 &
1\end{bmatrix},K_4=-\begin{bmatrix}2.32\cr
4.9\end{bmatrix}\]
\vspace{-0.2in}

 In all four cases the local observer convergence rates  are all~$2$.
 
\vspace{-0.1in}

This system was simulated with $x(0) = [3\;2\;4\;1]'$
as the initial state of the process,  $w_1(0) =[5\; 5]'$, $w_2(0) =[5\; 5]'$,
$w_3(0) = [5\; 5]'$, and $w_4(0)
=[5\; 5]'$ as the initial states of the
four local observers,  $x_1(0) = x_2(0)  
=[5\; 5\;5\;5]'$, and $ x_3(0) = x_4(0)
=[4\; 4\;4\;4]'$ as the initial estimates
of the four local estimators.

\begin{figure}[h]
\centerline{\includegraphics [ height
=2.2in]{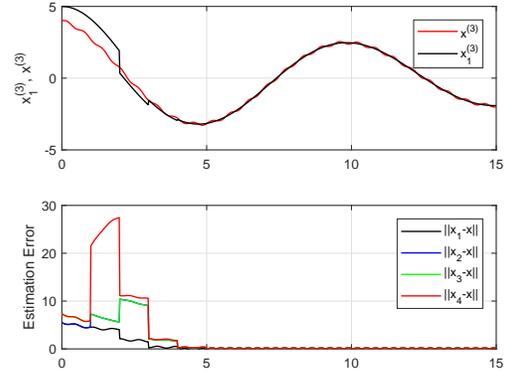}}
 \caption{Performance in the face of system noise}
   \label{zop1}
\end{figure}

\vspace{-0.1in}
Three simulations were performed. The first is intended to
demonstrate performance in the face of system noise. For this a
modified process dynamic  of the form $\dot{x} =Ax +b\nu$ is assumed
where $b= [1\; 1\;1\;1]'$ and  $\nu = \cos
10t$ is system noise. Traces of this simulation are shown in Figure
\ref{zop1} where $x_1^{(3)}$ and $x^{(3)}$ denote the third
components of $x_1$ and $x$ respectively.
 Only the trajectory of   $x_1^{(3)}$  is plotted because   for agent $1$ only the third component is unobservable, and all the other components are observable.
\vspace{-0.1in}
\vspace{-0.1in}

\begin{figure}[h]
  \centerline{\includegraphics [ height =2.2in]{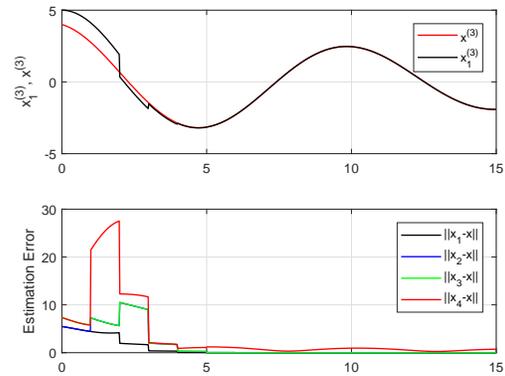}}
  \caption{Performance in the face of
a perturbed event time sequence}
   \label{zop2}
\end{figure}

The second simulation, which is without system noise, is intended to
demonstrate the hybrid observer's robustness against a small change
in the event time sequence of one of the agents. The change
considered presumes that the event times of agent 4 occur $.2T$ time
units before the  the event times of the other  three agents. Traces
of this simulation are shown in Figure \ref{zop2}.

\vspace{-0.1in}

\begin{figure}[h]
\centerline{\includegraphics [ height
=2.2in]{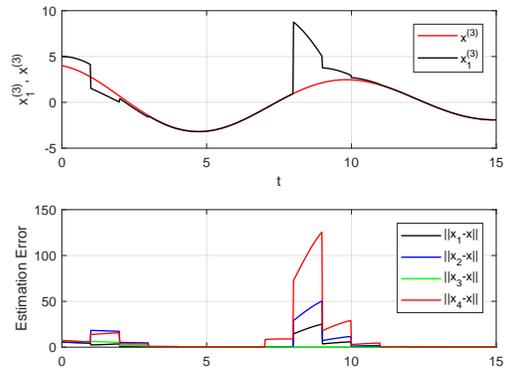}}
 \caption{Performance in the face
of abrupt node changes}
   \label{zop3}
\end{figure}

The third simulation, also without system noise,  is intended to
demonstrate the hybrid observer's resilience against  the
disappearance of agent 4 at time $t=5$ and also against agent $4$'s
re-emergence  at time $t=7$. Traces of this simulation are shown in
Figure \ref{zop3}.

Disruption appearing at the beginning of the traces for all three
simulations are due to initial conditions and are not important. In
the third simulation, the loss of agent 4 at $t=5$ does not appear
to have any impact whereas the trace shows that the re-emergence of
agent 4 at $t=7$ briefly effects performance. While the claims in
this paper do not consider the possibility of of agent re-emergence,
it is not surprising that this event does not cause misbehavior
because the time between the loss and gain of the agent, namely $2$
time units, is large compared to the time constants of the observer.
Clearly much more work needs to be done here to better understand
rapidly occurring and re-occurring losses and gains
 of agents.

\vspace{-0.1in}

\section{Concluding Remarks} \label{sec:conclusion}

\vspace{-0.1in}

One of the nice properties of the hybrid observer discussed in this
paper is that it is {\em resilient}. By this we mean that under
appropriate conditions it is   able to continue to provide
asymptotically correct estimates of $x$, even if communications
between some agents break down or if
  one or several of the agents
joins or leaves the network. The third simulation provides an
example of this capability. As pointed out earlier,
 further research is needed to more fully understand observer resilience,
 especially the situation when agents  join or leave the network.

Generally one would like to choose $T$ ``small'' to avoid
unnecessarily large error overshooting between event times.
Meanwhile it is obvious  from \rep{koop} that the larger the number
$p$ and consequently the number of iterations $q$ on each event-time
interval,  the faster the convergence. Two considerations limit the
value of $q$ - how fast the parameter estimators can compute and how
quickly information can be transmitted across the network. We doubt
the former consideration will prove very important in most
applications, since digital processors can be quite fast and the
computations required are not so taxing.  On the other hand,
transmission delays will almost certainly limit the choice of  $q$.
A model which explicitly takes such delays into account will be
presented in another paper.

A  practical issue is that the development in this paper does not
take into account measurement noise. On the other hand, the observer
provides exponential convergence and this suggests that if noisy
measurements are considered,  the observer's performance will
degrade gracefully with increasing noise levels. Of course one would
like an ``optimal'' estimator for such situations in the spirit of a
Kalman filter. Just how to formulate and solve such a problem  is a
significant issue for further research.

\vspace{-0.1in}
\begin{ack}                                
This work was supported by NSF grant 1607101.00, AFOSR grant FA9550-16-1-0290, and ARO grant  W911NF-17-1-0499.
\end{ack}
\vspace{-0.1in}

\bibliographystyle{unsrt}
\bibliography{observer2018,steve}




\end{document}